\documentstyle[12pt,aasms4]{article}
\slugcomment{to be published in: \quad {\bf  Feb. 2000, ApJ, 529} }
\lefthead{Garc\'{\i}a-Barreto and Moreno }
\righthead{Mass Model}

\begin{document} 

\hyphenation {non-re-la-ti-vi-stic}
\def\ab{a_{{\rm BR}}}
\def\abi{a_{{\rm BRI}}}
\def\pbi{p_{{\rm BRI}}}
\def\qbi{q_{{\rm BRI}}}
\def\abii{a_{{\rm BRII}}}
\def\pbii{p_{{\rm BRII}}}
\def\qbii{q_{{\rm BRII}}}
\def\rii{\rho_{{\rm BRII}}}
\def\gb{\gamma_{{\rm BR}}}
\def\xbi{\xi_{{\rm BRI}}}
\def\zbi{\zeta_{{\rm BRI}}}
\def\zbii{\zeta_{{\rm BRII}}}
\def\xbii{\xi_{{\rm BRII}}}
\def\fb{f_{{\rm BRI}}}
\def\am{a_{{\rm min}}}
\def\mb{\mu_{{\rm BRI}}}

\title{The Remarkable Central Structure of the Barred Galaxy NGC 1415}

\author{J. A. Garc\'{\i}a-Barreto\altaffilmark{1} and E. Moreno\altaffilmark{1}}
\altaffiltext{1}{Instituto de Astronom\'{\i}a, 
Universidad Nacional Aut\'onoma de M\'exico, Apdo Postal 70-264, M\'exico D.F. 
04510, M\'exico.
\\
tony@astroscu.unam.mx, edmundo@astroscu.unam.mx}

\begin{abstract} 
\noindent
  A remarkable structure is observed in the innermost regions of the 
barred galaxy NGC 1415 : a small stellar bar, bright circumnuclear 
ionized gas seen in H${\alpha}$, two bright ionized gas sources, seen in 
H${\alpha}$, just beyond the ends of the small bar and a boxy distribution
of optical continuum. We have developed a mass distribution model consisting in a disk, a bulge, and a bar to approximate the observed central morphology and its surface brightness. In order to reproduce the observed optical brightness distribution a two-component bar was used, with one component to model the elongated isophotes of the bar and the second component to model the boxy-shaped isophotes. We interpret the circumnuclear ionized gas as forming a circumnuclear ring and the two bright sources as a nuclear and/or circumnuclear outflow slightly out of the plane of the disk.

\keywords{galaxies: masses ---
galaxies: individual: NGC 1415 ---
galaxies: interstellar matter ---
galaxies: spiral
}
\end{abstract}
\section{Introduction}

Numerical N-body simulations of disk galaxies have shown that they may 
develop a high eccentricity bar with an elliptical shape (Martin 1995; Martin \& Friedli 1997). Sparke \& Sellwood (1987), Combes et al. (1990) and Athanassoula et al. (1990) have shown, however, that bars are more rectangular than elliptical. In our survey of H${\alpha}$ emission from barred galaxies  we included the galaxy NGC 1415 where its {\it I} emission shows elongated and boxy-shaped isophotes in the central regions (Garc\'{\i}a-Barreto et al. 1996).

NGC 1415 is classified as an SBa in the Revised Shapley Aames catalog (Sandage \& Tammann 
1987). Long exposure photographs of this galaxy, like those in the Hubble Atlas (Sandage 1962), indicate a long bar at a P.A. ${\simeq}130^{\circ}$ with two spiral arms 
originating from a central structure. The heliocentric velocity is :
V$_{HI}{\simeq}1553$ km s$^{-1}$ (Huchtmeier 1982), 
V$_{optical}=1566$ km s$^{-1}$ (Strauss et al 1992). Here we adopt a distance of D$=17.7$ Mpc (H$_{\circ}=75$ Kms$^{-1}$Mpc$^{-1}$; Tully 1988)
and therefore the linear scale is 1$''{\simeq}86$ pc. The inclination of the galaxy is about 65$^{\circ}$. More information about 
this galaxy can be found in Garc\'{\i}a-Barreto et al. (1996).

In this paper, we present a mass distribution model to reproduce the observed optical brightness distribution in the central region of the galaxy in the broad band filter {\it I}, with the goal of reproducing the observed boxy-shape isophotes. The models include analytic density functions for a disk, a two component bar and a small bulge. In $\S$ 2 we describe the spatial structure observed in the {\it I} band and
H${\alpha}$ images. The {\it I} image reveals the existence of a small bar with boxy isophotes in the central region while 
the H${\alpha}$ images reveal the existence of circumnuclear ionized gas 
plus two bright H${\alpha}$ sources straddling the nucleus at a slightly different position angle than the bar. Some characteristics of a SBa galaxy are reproduced in $\S$ 3. Our model for the mass 
distribution is presented in $\S$ 4. In $\S$ 5 we describe the input parameters and the results of the fits, and in $\S$ 6 we briefly discuss how well the model approximates the continuum observations and digress on the origin of the H$\alpha$ circumnuclear structure. Finally in $\S$ 7 we summarize the conclusions.

\section{ Spatial Distribution of the Optical Continuum and H$\alpha$ emissions in the central Region}

	Optical images with both a broadband and a narrowband filters were re-analyzed. Both images are taken from the survey work of barred spiral galaxies by Garc\'{\i}a-Barreto et al. (1996) where details about the observations and the data reduction can be found. The continuum image was obtained with a broadband filter, {\it I}, with ${\lambda}_c\simeq8040$ \AA, ${\Delta}{\lambda}\simeq1660$ \AA. The H$\alpha$+N[II] continuum-free image was obtained with a set of two narrowband filters with ${\lambda}{\simeq}~6459$ \AA  \ with ${\Delta}{\lambda}{\simeq}~101$ \AA \ and ${\lambda}{\simeq}~6607$ \AA  \ with ${\Delta}{\lambda}{\simeq}~89$ \AA. No image has been flux calibrated. Figures 1a and 1b show different continuum emission contours from the {\it I} filter, Fig. 2 shows the continuum-free H${\alpha}$ image and Fig. 3 shows the the H${\alpha}$ emission (grey scale) superposed to the continuum, {\it I}, emission (contours).

	Figures 1a and 1b show images of the innermost ${\sim}125''$ region of the barred galaxy NGC~1415 in the {\it I} filter. Four basic structures are 
identified, namely, (1) the compact nucleus, (2) an elongated stellar bar 
with a PA ${\sim}150^{\circ}$ and of ${\sim}18''$ in diameter, (3) the boxy 
contours outside the elongated bar, and (4) the starting points of the 
spiral arms. Figure 2 shows the image of the spatial distribution of ionized gas (H${\alpha}$ emission) . There are three characteristics 
worth emphasizing : (1) there is $no$ H${\alpha}$ emission 
from the compact nucleus, (2) there is an indication of a circumnuclear 
ring, and (3) there are two bright regions straddling the nuclear ring in 
the northwest and southeast with weak emission connecting to the latter. The position angle of a line joining the two bright spots is 
${\sim}166^{\circ}$. Figure 3 shows a superposition of the continuum and 
the continuum-free H${\alpha}$ images. 

	The H${\alpha}$ emission originates from 
the inner regions of the boxy distribution of the continuum emission, 
that is, within the inner $20''$. The P.A. of the two bright H${\alpha}$ regions is off by ${\sim}15^{\circ}$ from the P.A. of the elongated isophotes of the stellar bar. This suggests that each 
H${\alpha}$ bright region is trailing the ends of the inner stellar bar, assuming the inner bar is rotating clockwise as suggested by 
the starting points of the exterior spiral arms.

\subsection{Position of the Compact Nucleus of NGC 1415}

 Table 1 lists the position associated with NGC 
1415 reported by several authors, using different
techniques and different wavelengths. Gallou\"et, Heidmann \& Dampierre (1975) determined the optical position by fits of the 
optical emission from a Palomar Sky Survey film using a Zeiss micrometer machine. Fits of the optical emission from ESO plates were done by 
Loveday (1996). Despite the  high angular resolution used by Saikia et al. (1994) in their radio continuum measurements, these authors did not propose an improved position for the nucleus. Condon et al. (1990) find that the position of the peak of the radio emission is displaced from the mean position of the far infrared emission. Although there is a consensus about the brightness centroid position of the galaxy, there is a clear need to determine accurately the position of the dynamical center, namely, of the compact nucleus.

\subsection{The Inner Stellar Bar}

Figure 1b and Fig. 3 show the contour levels of the continuum emission 
from the central region of the galaxy. The stellar bar is prominent 
and it has an elliptical shape. Its semi-major axis dimension is 
${\sim}12.5''$ at a PA ${\sim}150^{\circ}$ on the plane of the sky as 
measured from pixel positions on the screen. Its semi-minor axis dimension is ${\sim}7.5''$ at $~10$\% of the relative intensity of the peak of the compact nucleus. 
This gives an axes ratio b/a$\sim$0.6. This stellar bar is small in length. Out of 137 galaxies (mostly from NGC catalog) where bar lengths were measured by Martin (1995), 101 have larger bar lengths than the one in NGC 1415; 76 out of 
81 galaxies listed by Kormendy (1979) have larger bar diameters and all of the NGC galaxies measured by  Chapelon, Contini \& Davoust (1999) have larger bar diameters than in NGC 1415. From 97 Markarian barred galaxies only 28 have larger angular dimensions of their bars than the one in NGC 1415, but their linear dimensions are probably larger since Markarian galaxies are far more distant than NGC galaxies (Chapelon, Contini \& Davoust 1999). The stellar bar, in NGC 1415, does not show any sign of a dust lane.

\subsection{The Boxy Isodensity Contours}

	Boxy-shaped isophotes are observed in the
continuum image of NGC 1415, Figure 1a. They start at a radius larger 
than the end of the elongated isophotes and maintain their shape until the start 
of the outer spiral arms, that is, up to a radius of ${\sim}24''$. The outer spiral arms start from the boxy-shaped isodensity contours 
in the northwest and in the southeast directions suggesting that the galaxy is 
rotating clockwise. These arms exist within a weaker continuum emission 
whose shape is elliptical on the plane of the sky, with length 
${\sim}104''$ at a PA${\sim}130^{\circ}$. 

	The overall optical appearance of NGC~1415 indicates a ``normal'' barred spiral galaxy, but the central region displays symmetrical gas structures with vigorous star forming activity. In $\S$ 4 different mass distribution models are discussed in order to approximate the 
surface brightness distribution of the central region and the boxy spatial structure. There is some evidence of weaker emission at larger radii from images of NGC 1415 in the POSS plate. Its interpretation lies beyond the scope of this paper.

\section {Characteristics of an SBa galaxy}

	In order to understand the morphology of NGC 1415, we include in this section a brief recollection of the characteristics of galaxies classified as SBa from Sandage (1962) and Kormendy (1979, 1981). A summary of the characteristics are : {\bf 1)} SBa galaxies are the earliest of the true barred spiral forms, {\bf 2)} the arms of an early SBa are poorly defined, smooth in texture, broad and fuzzy, {\bf 3)} the arms in a late SBa are relatively narrow and well developed and some show signs of being resolved, are tightly wound and form nearly circular arcs that appear in projection as elliptical loops, {\bf 4)} The spiral pattern can originate in two ways, either by emerging tangent to a complete internal ring at the edge of the lens, or by springing from the ends of the bar, {\bf 5)} the bars are devoid of dust, are smooth in texture and have no trace of resolution into knots or stars, {\bf 6)} lenses are found in 54\% of SBa galaxies but in none of late type (SBc), {\bf 7)} lenses tend to be triaxial, flattened ellipsoids, with a prefered equatorial axes ratio of $\simeq0.9$, {\bf 8)} half of SBa galaxies have bars and lenses of the same size, {\bf 9)} bulges are generally triaxial elongated perpendicular to the bar and in some cases elongated into a secondary nuclear bar, {\bf 10)} bulges rotate very rapidly, {\bf 11)} there is probably an ILR in the outer part of the bulge (Kormendy 1981).

	Observations of NGC 1415 clearly indicate smooth spiral arms starting from a boxy structure. The elongated bar shows a smooth texture and no signs of a dust lane or any individual knots. There is no clear evidence of a lens in NGC 1415. These characteristics fully guaranty the classification as an SBa galaxy, except that,as will be seen in $\S$ 5.2,  in order to reproduce the brightness distribution in the innermost regions and the boxy ishophotes our model only requires a small size spherically symmetric bulge rather than a large one.

\section {Mass Distribution Model}

	In order to reproduce the morphology of barred  galaxies, numerical simulations of Sparke \& Seelwood (1987), Combes et al. (1990) and Athanassoula et al. (1990), employ a two-component initial distribution of particles, namely, a Toomre's disk with a given surface density and a spherically symmetric Plummer bulge-like component with its corresponding volume density.

	In this section we describe our approach. We model a barred spiral galaxy using analytic functions consiting of three components, namely,  a disk-like component, a spherical bulge, and a bar. 

\subsection{Components}

To model the observed surface brightness in NGC 1415, three components are
considered for its mass distribution: disk, bulge, and bar. The disk is 
represented with a classic oblate spheroid with similar strata (i.e. the
isodensity surfaces are aligned, concentric spheroids, all with the same
axial ratio as that of the boundary surface). Its density is given by the classic form (Schmidt 1956):

\begin{equation}
\rho_D(a) = \frac{q_D}{a} + p_D a \qquad , \qquad a \leq a_D
\end{equation}

\noindent where $q_D$, $p_D$ are constants, $a$ is the major semi-axis of a 
similar spheroidal
surface within the oblate spheroid, and $a_D$ is the equatorial radius of this
spheroid. At the boundary $a = a_D$ the density of the disk is zero, so 
that $q_D = -p_D a_D^2$.

The bulge is considered spherically symmetric with a density function of the form given by Eq. (1). Instead of $a$, however, the distance variable is now r, the distance to the galactic center. The constants are $q_B$, $p_B$, and the radius of the bulge $r_B$, at which point its density goes to zero.

We found that a satisfactory fit to the observed optical brightness distribution in NGC 1415 requires a two-component bar. The bar in this case has a triaxial mass distribution, and its two components are: 1) an ellipsoidal mass distribution with
similar strata, with density of the form given by Eq. (1) where the variable $a$ is the major semi-axis of an internal similar ellipsoid; the associated constants are $\qbi$, $\pbi$, $\abi$ (the density of this first component drops to zero at the boundary $a = \abi$),  and  2) a triaxial mass distribution whose isodensity surfaces are of the form

\begin{equation}
\left( \frac{x}{a}\right)^4 + \left( \frac{y}{b}\right)^4 +
\left( \frac{z}{c}\right)^4 = 1 ,
\end{equation}

\noindent with constant ratios $b/a$, $c/a$ (i.e. similar surfaces) and where $x$, $y$, and $z$ are cartesian coordinates.
Eq. (2)
is a special case of a generalized ``ellipsoidal" shape for a mass
distribution. Athanassoula et al. (1990) have considered the generalized
``ellipse" to fit boxy isophotes in galaxies. The axes of the surfaces given
by Eq. (2) are aligned with those of the first ellipsoidal component, the
major axes pointing along the $x$ direction on the equatorial plane of the 
disk. The major semi-axis of this second component is $\abii$. 
The reason to introduce this second component is the need to model the
boxy isophotes observed in Fig. 1: by itself the projection on the plane of
the sky of isodensity surfaces given by Eq. (2) produces isophotes of this
type.

A decreasing density function in the second component of the bar can take one of the three following forms ($a\leq\abii$):

\begin{eqnarray}
\rii (a) & = & \frac{\qbii}{a} + \pbii a \qquad ,  \nonumber\\
\rii (a) & = & \qbii + \pbii a \qquad ,  \\
\rii (a) & = & \qbii + \pbii a^4 \qquad . \nonumber
\end{eqnarray}

The first form is the same as in the ellipsoidal component. Note that $\qbii$ and 
$\pbii$ are different for the three forms, while $\rii(\abii) = 0$ in all cases.

\subsection{Surface Brightness}

Each component of the mass distribution (disk, bulge and bar) is considered to have its own
position-independent mass-to-light ratio $\gamma$. Both components of the bar have the same $\gb$ (the mass-to-light ratio of the bar) to keep the number of free parameters to a minimum.
In Fig. 4 we show the three components of the system, and the line of
sight with angular coordinates $(\phi, \theta)$. The Cartesian axes $x'$, 
$y'$, $z'$
are oriented as follows: the $x'$ axis is perpendicular to the azimuthal
plane defined by the axes $z$, $z'$, while $x'$, $y'$ axes define the plane of the sky. The centers of coordinates of both Cartesian systems $(x,y,z)$ and
$(x',y',z')$ coincide, being located at the galactic center. The inset only shows the center of the $(x',y',z')$ system displaced for the sake of clarity.

The surface brightness of an ellipsoidal mass distribution with similar
strata can be derived from the equations given by Stark (1977) and Binney
(1985). In this case, the surface brightness of the disk (oblate spheroid), the bulge,
and the first component of the bar, can be readily found. The main equations
can be summarized as follows:

\noindent We call $\xi_D = (c/a)_D = c_D/a_D$, with $c_D < a_D$, the semi-axes ratio of the spheroid representing the disk. Similarly we define $\zbi = (b/a)_{{\rm BRI}}$, $\xbi = (c/a)_{{\rm BRI}}$, $a \geq b \geq c$, representing the first component of the bar. For a given line of sight with coordinates $(\phi,\theta)$ (see Fig. 4), the projection of the disk's boundary on the plane of the sky $(x',y')$, delimit a region specified by

\begin{equation}
a_{{\rm minD}}(x',y') = \left( x'^2 + \frac{y'^2}{f_D\xi_D^2}\right)^{1/2}
\leq a_D ,
\end{equation}

\noindent with $f_D\xi_D^2 = \xi_D^2\sin^2\theta + \cos^2\theta$. Similarly, for the bulge this region is

\begin{equation}
a_{{\rm minB}}(x',y') = (x'^2 + y'^2)^{1/2} \leq r_B ,
\end{equation}

\noindent and for the first component of the bar:

\begin{equation}
a_{{\rm minBRI}} (x',y') = (\Omega_1 x'^2 + \Omega_2 y'^2 + \Omega_3 
x'y')^{1/2} \leq \abi ,
\end{equation}

\noindent with

\begin{eqnarray}
\Omega_1 & = & \frac{\xbi^2\sin^2\theta + (\zbi^2\sin^2\phi + cos^2\phi)
               \cos^2\theta}{\fb\zbi^2\xbi^2}\nonumber \\
\Omega_2 & = & \frac{\zbi^2 cos^2\phi + \sin^2 \phi}{\fb \zbi^2 \xbi^2}
               \nonumber \\
\Omega_3 & = & \frac{(1 - \zbi^2)\cos\theta\sin(2\phi)}{\fb\zbi^2 \xbi^2}
               \nonumber \\
\fb\zbi^2\xbi^2 & = & \zbi^2\cos^2\theta + \xbi^2(\zbi^2\cos^2\phi + \sin^2\phi)
               \sin^2\theta . \nonumber
\end{eqnarray}

The surface brightness of each component at a point $(x',y')$ within the
corresponding region is

\begin{eqnarray}
B_D(x',y') & = & \frac{q_D}{2\pi\sqrt{f_D}\gamma_D} F(\mu_D(x',y')) \qquad ,\\
B_B(x',y') & = & \frac{q_B}{2\pi\gamma_B} F(\mu_B(x',y')) \qquad ,\\
B_{{\rm BRI}}(x',y') & = & \frac{q_{{\rm BRI}}}{2\pi\sqrt{f_{{\rm BRI}}}
\gamma_{{\rm BR}}} F(\mu_{{\rm BRI}}(x',y')) \qquad ,
\end{eqnarray}

\noindent with $\mu_D(x',y') = a_{{\rm minD}}(x',y')/a_D$, $\mu_B(x',y') =
a_{{\rm minB}}(x',y')/r_B$,
$\mb(x',y') = a_{{\rm minBRI}}(x',y')/\abi$, and $F(\mu) = \left(
1 - \frac{1}{2}\mu^2\right)\ln \frac{1 + \sqrt{1 - \mu^2}}{\mu} -
\frac{1}{2}\sqrt{1 - \mu^2}$. The coefficients $q_D$, $q_B$, and $\qbi$
can be expressed in terms of the total mass of the corresponding component: 
$q_D= M_D/(\pi\xi_D a_D^2)$, $q_B = M_B/\pi r_B^2$, $\qbi = M_{{\rm BRI}}/(\pi\zbi\xbi\abi^2)$. The total masses of the disk, bulge, and first component of the bar are $M_D$, $M_B$, and $M_{{\rm BRI}}$ respectively.

In terms of the surface brightness of the disk at $\mu_D = 1/2$: 
$B_D(\mu_D =1/2) = M_Dk_0/2\pi^2\sqrt{f_D}\gamma_D\xi_D a_D^2$,
$k_0$= F(1/2), i.e. on the disk's isophote at the middle of the elliptic region given by Eq. (4), the surface brightnesses in Eqs. (7), (8), and (9) can be written as:

\begin{eqnarray}
\frac{B_D(x',y')}{B_D\left(\mu_D = \frac{1}{2}\right)} & = & 
\frac{F\left(\mu_D (x',y')\right)}{k_0} \\
\frac{B_B(x',y')}{B_D\left(\mu_D = \frac{1}{2}\right)} & = &
\frac{\xi_D\sqrt{f_D}}{k_0}
\left(\frac{M_B}{M_D}\right) \left(\frac{a_D}{r_B}\right)^2
\frac{\gamma_D}{\gamma_B}F\left(\mu_B(x',y')\right)\\
\frac{B_{{\rm BRI}}(x',y')}{B_D\left(\mu_D = \frac{1}{2}\right)} & = &
\frac{\xi_D}{k_0\zbi\xbi} \left( \frac{f_D}{\fb}\right)^{1/2} \left( \frac{M_{{\rm BRI}}}{M_D}\right)\left( \frac{a_D}{\abi}\right)^2 \frac{\gamma_D}{\gamma_{{\rm BR}}} F\left(\mu_{{\rm BRI}}(x',y')\right)
\end{eqnarray}

The expression for the surface brightness of the second component of the
bar is a little more complicated. This brightness at a point $(x',y')$ inside
its projected region is

\begin{equation}
B_{{\rm BRII}}(x',y') = \frac{1}{4\pi\gamma_{{\rm BR}}} \int_{z'_1}^{z'_2}
\rii(a)dz'
\end{equation}

\noindent where $z'_1$, $z'_2$ $(z'_1 \leq z_2')$ are the values of $z'$ (see Fig. 4) at the two intersection points of the line of sight through the point $(x',y')$, with the boundary surface of this second component. The density $\rii(a)$ is one of the three forms given in Eq. (3). The semi-axes ratios in the second component are $\zbii = (b/a)_{{\rm BRII}}$, $\xbii = (c/a)_{{\rm BRII}}$. 

The function $a(z')$ in Eq. (13) is given by

\begin{equation}
a^4(z')  = A_1z'^4 + A_2z'^3 + A_3z'^2 + A_4z' + A_5
\end{equation}

\noindent where $A_1,\dots, A_5$ are the following functions:
\begin{eqnarray}
A_1 & = & \alpha_1^2 + \frac{\beta_1^2}{\zbii^4} + \frac{\gamma_1^2}{\xbii^4}
          \nonumber \\
A_2 & = & 2\left(\alpha_1\alpha_2 + \frac{\beta_1\beta_2}{\zbii^4} +
          \frac{\gamma_1\gamma_2}{\xbii^4}\right) \nonumber \\
A_3 & = & \alpha_2^2 + 2\alpha_1\alpha_3 + \frac{\beta_2^2 + 2\beta_1\beta_3}
          {\zbii^4} + \frac{\gamma_2^2 + 2\gamma_1\gamma_3}{\xbii^4}\nonumber\\
A_4 & = & 2\left(\alpha_2\alpha_3 + \frac{\beta_2\beta_3}{\zbii^4} + 
          \frac{\gamma_2\gamma_3}{\xbii^4}\right)\nonumber \\
A_5 & = & \alpha_3^2 + \frac{\beta_3^2}{\zbii^4} + \frac{\gamma_3^2}{\xbii^4}
\nonumber
\end{eqnarray}

\noindent and

\begin{eqnarray}
\alpha_1 & = & \sin^2\theta\cos^2\phi \nonumber \\
\alpha_2 & = & -x'\sin\theta\sin(2\phi) + y'\cos^2\phi \sin(2\theta)\nonumber\\
\alpha_3 & = & x'^2\sin^2\phi + y'^2\cos^2\theta\cos^2\phi - x'y'\cos\theta
               \sin(2\phi)\nonumber \\
\beta_1 &  = & \sin^2\theta\sin^2\phi \nonumber \\
\beta_2 &  = & x'\sin\theta\sin(2\phi) + y'\sin^2\phi \sin(2\theta) \nonumber\\
\beta_3 &  = & x'^2\cos^2\phi + y'^2\cos^2\theta\sin^2\phi + x'y'
               \cos\theta\sin(2\phi) \nonumber \\
\gamma_1 & = & \cos^2\theta \nonumber \\
\gamma_2 & = & -y'\sin(2\theta) \nonumber \\
\gamma_3 & = & y'^2\sin^2\theta\nonumber
\end{eqnarray}

The limits $z'_1$, $z'_2$ $(z_1' \leq z_2')$ in Eq. (13) are the two real 
roots of
$\abii^4 = A_1 z'^4 + A_2z'^3 + A_3z'^2 + A_4z' + A_5$. Given a point  $(x',y')$ on the plane of the sky, this point will be outside the region defined by
the projection of the second component of the bar if the four roots of
this equation are complex. The integration of Eq. (13) is straight forward  in the case where the density $\rii(a)$ is expressed as the third form in Eq. (3). This was the main reason for considering this form. Numerical integration was used for the other two density forms.

The expressions for the surface brightness
of the second component of the bar are the following:

\begin{itemize}
\item when considering a density function $\rii(a) = \frac{\qbii}{a} + \pbii a$:

\begin{equation}
\frac{B_{{\rm BRII}}(x',y')}{B_D\left(\mu_D = \frac{1}{2}\right)} = 
\frac{\pi\xi_D\sqrt{f_D}}{12\zbii\xbii k_0k_1}\left(\frac{M_{{\rm BRII}}}
{M_D}\right)\left(\frac{a_D}{\abii}\right)^2\frac{\gamma_D}{\gamma_{{\rm BR}}}
S_1 ,
\end{equation}

\item when considering a density function $\rii(a) = \qbii + \pbii a$:

\begin{equation}
\frac{B_{{\rm BRII}}(x',y')}{B_D\left(\mu_D = \frac{1}{2}\right)} = 
\frac{\pi\xi_D\sqrt{f_D}}{4\zbii\xbii k_0k_1}\left(\frac{M_{{\rm BRII}}}
{M_D}\right)\frac{1}{\abii}\left(\frac{a_D}{\abii}\right)^2\frac{\gamma_D}
{\gamma_{{\rm BR}}}S_2 ,
\end{equation}

\item when considering a density function $\rii(a) = \qbii + \pbii a^4$:

\begin{equation}
\frac{B_{{\rm BRII}}(x',y')}{B_D\left(\mu_D = \frac{1}{2}\right)} = 
\frac{7\pi\xi_D\sqrt{f_D}}{64\zbii\xbii k_0k_1}\left(\frac{M_{{\rm BRII}}}
{M_D}\right)\frac{1}{\abii}\left(\frac{a_D}{\abii}\right)^2\frac{\gamma_D}
{\gamma_{{\rm BR}}}S_3 ,
\end{equation}

\end{itemize}

\noindent $k_1 \simeq 0.81025$ is a constant, and $S_1,S_2,S_3$ are the following integrals 
\begin{eqnarray}
S_1 & = & \int_{z'_1}^{z'_2} \frac{1}{a(z')}\left\{ 1 - \left[\frac{a(z')}
          {\abii}\right]^2\right\}dz' \nonumber \\
S_2 & = & \int_{z'_1}^{z'_2} \left\{ 1 - \frac{a(z')}{\abii}\right\}dz'\nonumber\\
S_3 & = & \int_{z'_1}^{z'_2}\left\{ 1 - \left[\frac{a(z')}
          {\abii}\right]^4\right\}dz'\nonumber
\end{eqnarray}

By summing Eqs. (10), (11), (12), with either of Eqs. (15), (16), (17), 
we derive the total surface brightness $B_{{\rm TOT}}(x',y')/B_D(\mu_D = 1/2)$ at any point 
$(x',y')$ on the plane of the sky. Of course, we must always check if this point lies
inside a projected region, in order to ensure that the contribution to the total brightness
belong to the appropiate component of the mass distribution. The isophotes corresponding to $B_{{\rm TOT}}(x',y') = $ {\it constant} are found numerically, once all pertinent data is
introduced in the equations. In the following section, we describe such a  procedure for finding the best fit which approximates the optical surface brightness distribution of the central region of NGC 1415.

\section{Model Results}

\subsection{Input Data}

The following quantities are considered as input data to our model : a visual inspection of the {\it I} band image of NGC 1415 suggests a value of $a_D = 9$~ kpc as a representative radius of the disk, assuming a ratio of minor to major axes of $\xi_D  = 0.1$. Isophotes along the major and minor axis of the bar suggest an axial ratio $\sim 0.6$, therefore in 
a first try we take $\zbi = \xbi = \zbii = \xbii = 0.6$. The outermost boxy
contour in Fig. 1 gives an estimate for the major semi-axis of the bar of between 1.3 and 1.5 kpc; therefore we first assume $\abi =  \abii = 1.5$~kpc, but as we point below, the equality between $\abi$ and $\abii$ must be dropped in order to improve the fit. We obtain that $\abi$ is slightly less than $\abii$. The radius of the bulge is unknown at the start, and can be treated as a variable parameter for improving the fit. Other variables are the mass-to-light ratios and the ratios of masses of the mass distribution components appearing in Eqs. (11),(12),(15),(16), and (17).

The inclination of the galaxy with respect to the line of sight is reported to be $\theta = 65^\circ$. An inspection of Fig. 1
suggests that $\phi$ is close to $90^\circ$. In the fit, we adopt $\phi = 70^\circ$. As input data for surface brightness to our model, we have taken some representative lines across the observed {\it I} image. These are the following: $x' = 0.0$, $\pm 0.2$, $\pm 0.4$, $\pm 0.8$,
$\pm 1.2$, $\pm 1.4$~kpc; $y' = 0.0$, $\pm 0.2$, $\pm 0.4$, $\pm 0.6$, 
$\pm 0.8$~kpc. The line $y' = 0$ is the major axis of
the elliptic projected region of the disk while $x' = 0$ is the corresponding line of the minor axis. In the left panels of Figs. 5 and 6,  we show the observed surface brightness along some $x'$  and $y'$ lines.
   
The surface brightness given by our model has a logaritmic singularity at
$(x',y') = (0,0)$. In order to maintain a finite brightness at the center,  we have considered that the density given by Eq. (1), which is used in three components (disk, bulge and bar) of the model, should only apply beyond $a \geq a_{{\rm L}}$, where $a_{{\rm L}}$ is a fixed distance. We have therefore adopted a constant density, $\rho(a_L)$, for $a \leq a_{{\rm L}}$. A value $a_{{\rm L}}\sim 40$~pc has been found to be convenient to the fitting procedure.

\subsection{Model vs. Observation}

Some preliminary fits with $\abi =  \abii$ reproduced acceptably the form of the iso-contours shown in Fig. 1, but failed to give a good match to the observed surface brightness. Thus it was necessary to take different dimensions of the two components of the bar. The right panels of Figs. 5 and 6 show the brightness of our best fit along the $x'$  or $y'$  line whose corresponding observed brightness is given at the left panel. Two curves are given in these right panels: a fit with $\abi = 1$~kpc, $\abii = 1.3$~kpc shown with the continuous line, and a second fit with $\abi = 1$~kpc, $\abii = 1.5$~kpc shown with the dotted line. In both fits we considered the three forms given by Eq. (3) for the density of the second component of the bar, obtaining almost the same result. In Figs. 5 and 6 the curves of the model correspond to the fit when using the third form for the density in Eq. (3).

The upper panels in Fig. 7 show the isophotes in both fits. Compare these
fits with Fig. 1. The first panel shows the isophotes obtained in the first fit, with each form for the density in the second component of the bar. The third form in Eq. (3) gives the external curve in a given isophote; the first form gives the corresponding inner curve. The lower panels in Fig. 7 show the isophotes in the second fit with a line-of-sight coordinates $(\phi, \theta) = (90^\circ,90^\circ)$, but with $\xi_{{\rm D}} = 0.05$ to see more clearly the boxy structure.

The following quantities are obtained with the fits, using the third form of the density for the second component of the bar: 
\begin{itemize}
\item bulge radius $r_B = 0.3$~kpc, 
\item $(M_B/M_D)(\gamma_D/\gamma_B) = 0.00904$, 
\item $(M_{{\rm BRI}}/M_D)(\gamma_D/\gamma_{{\rm BR}}) = 0.152$, 
\item $(M_{{\rm BRII}}/M_D)(\gamma_D/\gamma_{{\rm BR}}) = 0.04$, $0.064$ in the first and second fit, respectively.
\end{itemize}

	The ratio of the luminosity of the bulge to that of the disk is $L_B/L_D~\sim~0.01$. The mass ratio $M_{{\rm BRI}}/M_{{\rm BRII}}=3.8$, while the luminosity ratio $L_{{\rm BR}}/L_D=0.192$ for the first fit and $M_{{\rm BRI}}/M_{{\rm BRII}}=2.37$, $L_{{\rm BR}}/L_D=0.216$ for the second fit. If $\gamma_D\sim\gamma_{{\rm BR}}\sim\gamma_B$, this would imply $M_B/M_D\sim0.01$ and $M_{{\rm BR}}/M_D\sim0.2$. Further kinematic data in NGC 1415 is needed to clarify this last issue (ie. rotation curve).

\section{Discussion}

\subsection{Our Mass Distribution Model}

	Our model using analytic functions for the disk, bulge and bar reproduces well the observed surface brightness distribution in NGC 1415, as can be judged in Figs. 5 and 6. Our two fits differ only in the major axis dimension of the two components of the bar. The first fit has $a_{{\rm BRII}}=1.3$ kpc while the second fit has $a_{{\rm BRII}}=1.5$ kpc. In both fits $a_{{\rm BRI}}=1$ kpc. Figures 5 and 6 show that both fits are comparable. 
Considering the observed dimensions of the bar in NGC 1415 as seen in Fig. 1 we favor the second fit to be the best model (see Fig. 7b).

	It is to be noticed that both fits of our model reproduce well the smooth observed transition of the isophotes of the bar: from the boxy-shape at large radii to elliptical isophotes at smaller radii to 
nearly circular isophotes at the very center of the galaxy. 

	From our model we deduce that the relative difference of position angles of the major axes of disk and bar is purely geometric and it is not necessarily due to a triaxiality of the bar with a variable axes ratio.

	Based on our results we suggest that a mass distribution with iso-density surfaces expressed as in Eq. 2 may be important to represent the central mass distribution in some barred galaxies. As a comparison  Athanassoula et al. (1990) only use a generalized ellipse equation to fit bar isophotes of observed galaxies, and they do not analyze the required iso-density surfaces to obtain boxy-shaped isophotes.

\subsection{Is the Bar driving the Spiral Arms?}

	Figures 1a and 1b show the spiral arms starting from NW and 
SE of the exterior boxy iso-density contours of the bar. Figure 6 of 
Sparke \& Sellwood (1987) shows a similar image from their N-body simulation,
that is, the spiral arms not starting exactly from the ends of the 
bar but at a different PA. In the case of NGC 1415, Figs. 1a and 1b
suggest that if the bar is driving the spiral arms, the bar must
be rotating clockwise such that the bar trails the start points of 
the spiral arms. Figure 3 shows, however, the two bright H${\alpha}$ 
regions just trailing the tips of the bar. Could it be that the inner stellar bar is rotating counter-clockwise? Without kinematical 
information our discussion is only speculative. There is a need to 
determine the velocity of the ionized gas in the central structure 
in order to make quantitative statements.

\subsection{The Circumnuclear ring and the two Bright H${\alpha}$ regions}

	It is well accepted by now that gas in a non-axisymmetric bar potential loses angular momentum and gas inside the co-rotation radius 
is transfered inwards while gas outside co-rotation is transfered outwards (Lynden-Bell \& Kalnajs 1972; Lynden-Bell \& Pringle 1974; Schwarz 1984; Combes \& Gerin 1985; Binney \& Tremaine 1987). The weak ionized gas emission just outside of the nucleus may be interpreted as a cirumnuclear ring forming near an inner Lindblad Resonance (Schwarz 1984). The western circumnuclear structure lies at a radius of 
$4''$ which corresponds to a distance of about 345 pc. Figure 3 shows 
the H${\alpha}$ emission from the central regions of NGC 1415. Notice 
that there is {\bf no} H${\alpha}$ emission from the compact nucleus. 
There is, however, weak emission connecting the circumnuclear ring with 
each bright spots further out. The SE H${\alpha}$ region is stronger 
than the NW region. 

	A very simple comparison of the stellar (optical continuum) and the warm ionized gas distributions (H$\alpha$) can be made in order to interpret the observations. It is to be noted the curved continuum contours to the SSE and SSW and SW of the compact nucleus as if indicating lack of stars in those locations. H$\alpha$ distribution indicates the presence of the bright SE spot and the SW circumnuclear arc. One interpretation is that the warm ionized gas is in front of the radiation from the stars, that is, it is slightly above the plane of the disk absorbing the stellar radiation. The ionized gas, then, would be in an inclined disk different from the inclination of the larger disk of stars.  Another interpretation (less likely) is that both the stars and ionized gas are on the same plane and the curved contours of continuum radiation would indicate different orbits caused somehow by the gravitational potential. 

	Each bright H$\alpha$ spot lies at slighty different distance from the compact nucleus. The NW bright region lies at a distance of ${\sim}10.5''$ corresponding to a linear distance from the nucleus of ${\sim}900$ pc. The SE bright region lies at a distance of ${\sim}8.5''$ corresponding to a linear distance from the nucleus of ${\sim}730$ pc. The two bright spots could themselves be in different planes from the circumnuclear ring. 

	In one interpretation the bright regions represent 
regions of density enhancement with a by-product of a burst of star 
formation where gas on disk orbits loses angular momentum and meets 
orbits trapped in the bar. Gas continues losing angular momentum 
and is seen along the bar until it reaches {\it x}$_2$ orbits and 
forms a circumnuclear structure seen more easily on the western side of the nucleus than on the eastern side.  Radio continuum emission has been 
detected from the central region of NGC 1415 by Condon et al. (1990).
Their map shows a bright central emission with extensions in the 
northwest - southeast direction covering approximately the same area as
our H${\alpha}$ image. However their angular resolution was 
not enough to resolve the emission from the compact nucleus and neither from 
any circumnuclear structure nor from the two bright H${\alpha}$ regions.

	Another interpretation of the two bright H${\alpha}$ regions is that they represent a kind 
of bipolar emission from the nucleus and or circumnuclear ring. The two H$\alpha$ bright sources would be slightly inclined with respect to the plane of the stellar disk. Similar morphology has been observed in other galaxies as NGC 1068 (Bland-Hawthorn et al. 1997) and NGC 3367 (Grac\'{\i}a-Barreto et al. 1998). The radio continuum radiation should be found polarized only in the SE region since the  emission from the NW would be depolarized as it passes through the warm interstellar matter in the disk. Observationally there is no H$\alpha$ emission from the compact nucleus (see Fig. 3) and it is difficult to say if there is any radio continuum emission, due to poor angular resolution in the maps of Condon et al. (1990). 

	Further observations are needed to find out where the bright H$\alpha$ sources lie: either on the plane of the sky or slightly inclined. A radio continuum map with higher angular resolution and polarization analysis would be necessary in order to study this. Similarly, HI and CO interferometric observations would be extremely useful in order to determine the atomic and molecular gas distributions. Additionaly an optical rotation curve with Fabry Perot techniques would be extremely useful in order to determine the dynamical center of this galaxy.

\section {Conclusions}
We have re-analyzed broadband and narrow band imaging observations of the central region of the barred spiral galaxy NGC~1415 from Garc\'{\i}a-Barreto et al. (1996). We have constructed a model for the mass distribution in the central region of NGC 1415 in order to approximate the observed optical surface brightness distribution. Our model includes a disk, a spherically symmetric bulge ($r_B=300$ pc) and a two-component bar: one component to reproduce the elongated isophotes and a second component to reproduce the boxy-shaped isophotes. The ratio of the masses of the first component to the second component is 2.37, a bar to disk luminosity ratio of 0.216, and a bulge to disk luminosity ratio of 0.01. If the mass to luminosity ratio is considered the same in the disk, bulge and bar, these results imply a ratio of masses of bulge to disk of 0.01 and bar to disk of 0.2.

	Our results suggest that a mass distribution with iso-density surfaces expressed as in Eq. (2) may be important to represent the central mass distribution in some barred galaxies.

Finally, we interpret the 
H${\alpha}$ emission as originating from a bipolar nuclear outflow  
slightly inclined with respect to the plane of the disk, although new observations are needed to confirm it.

\section*{Acknowledgements}
We would like to thank Luc Binette for his comments to improve the 
english version. 

\clearpage
\newpage
{\center \section*{References}}
\setlength{\parindent}{-1.0\parindent}

\ 

Athanassoula, E., Morin, S., Wozniak, H., Puy, D., Pierce, M.J., Lombard, J. \& Bosma, A. 1990, MNRAS, 245, 130

Binney, J. 1985, MNRAS, 212, 767

Binney, J., \& Tremaine, S. 1987 {\it Galactic Dynamics} Princeton Series in Astrophysics, Princeton, New Jersey.

Bland-Hawthorn, J., Lumsden, S.L., Voit, G.M., Cecil, G.N. \& Weisheit, J.C. 1997, ApSSc, 248, 177

Chapelon, S., Contini, T. \& Davoust, E. 1999, \aap, 345, 81

Combes, F., Debbasch, F., Friedli, D., \& Pfenniger, D. 1990, A\&A, 233, 82

Combes, F., \& Gerin, M. 1985, A\&A, 150, 327

Condon, J.J., Helou, G., Sanders, D.B., \& Soifer, B.T. 1990 ApJS, 73, 359

Gallou\"et, L., Heidmann, N., \& Dampierre, F. 1975 A\&ASS, 19, 1

Garc\'{\i}a-Barreto, J.A., Rudnick, L., Franco, F., \& Martos, M. 1998, AJ, 116, 111

Garc\'{\i}a-Barreto, J.A., Franco, F., Carrillo, R., Venegas, S. \& 
Escalante-Ram\'{\i}rez, B. 1996, Rev.Mexicana Astron. Astrofis., 32, 89

Huchtmeier, W.K. 1982, A\&A, 110, 121

Kormendy, J. 1979, ApJ, 227, 714

Kormendy, J. 1981, in The Structure and Evolution of Normal Galaxies, Ed. S.M. Fall \& D. Lynden-Bell (Cambridge: Cambridge Univ. Press)

Loveday, J. 1996, MNRAS 278, 1025

Lynden-Bell, D., \& Kalnajs, A.J. 1972, MNRAS, 157, 1

Lynden-Bell, D., \& Pringle, J.E., 1974, MNRAS 168, 603

Martin, P. 1995, AJ, 109, 2428

Martin, P. \& Friedli, D. 1997, A\&A, 326, 449

Saikia, D.J., Pedlar, A., Unger, S.W., \& Axon, D.J. 1994, MNRAS, 270, 46

Sandage, A. 1962 {\it The Hubble Atlas of Galaxies}, Carnegie Instituion of Washington, Washington, D.C.

Sandage, A., \& Tammann, G.A., 1987, {\it A Revised Shapley-Ames Catalog
 of Bright Galaxies}, Carnegie Institution of Washington, Washington, D.C.

Schwarz, M.P. 1984, MNRAS, 209, 93

Schmidt, M. 1956, B.A.N., 13, 15

Sparke, L.S., \& Sellwood, J.A. 1987, MNRAS, 225, 653

Stark, A. 1977, ApJ, 213, 368

Strauss, M.A., Huchra, J.P., Davis., M., Yahil, A., Fisher, K.B. \& Tonry, J. 1992, ApJS, 83, 29

Tully, R.B. 1988 {\it Nearby Galaxies Catalog}, (Cambridge Univ. Press, Cambridge)

\clearpage
\newpage

\section*{Tables}

%Table 1

\small 
\begin{table}
\caption[ ]{
Central Position of NGC 1415
}
\begin{flushleft}
\begin{tabular}{cccl}
\hline
\\
R.A.(1950.0)	&\ Decl.(1950.0)		& Method 
		& Reference \cr
\hline
\\ 
$3^h~~38^m~~46.0^s$		& $-22^{\circ}~~43'~~24.0''$ & optical  
		& Saikia et al. 1994 \cr
\\
$3^h~~38^m~~45.7^s$		& $-22^{\circ}~~43'~~19.3''$ & optical  
		& Loveday 1996 \cr
\\
$3^h~~38^m~~46.0^s$		& $-22^{\circ}~~43'~~25.0''$ & optical  
		& Gallou\"et, Heidmann \& Dampierre  1975 \cr
\\
$3^h~~38^m~~45.6^s$		& $-22^{\circ}~~43'~~30.0''$ & IRAS  
		& Condon et al. 1990 \cr
\\
\hline  
\end{tabular}
\end{flushleft}
\end{table}

\normalsize 
\clearpage
\newpage

\section*{Figure Captions}

\ 

FIGURE 1
a) Image in the broadband filter {\it I} (${\lambda}_{central} \simeq~8040$ \AA ); North is up, East is left, taken from Garc\'{\i}a-Barreto et al. (1996). Contours are in arbitrary units proportional to the peak flux. It was assumed that the optical position determined for the galaxy (ie. Gallou\"et, Heidmann \& Dampierre 1975) corresponds to the position of the compact nucleus. b) {\it I} iso-contour emission from the inner regions of NGC 1415. Notice 
the inner, elongated isophotes of the stellar bar and the contours just 
outside the bar forming a boxy shape. The spiral arms start from the northwest and southeast. The outer major axis is at PA$~{\sim}130^{\circ}$, while the inner at a PA~${\sim}150^{\circ}$.

FIGURE 2
Image of NGC~1415 in the narrowband filter H${\alpha}+NII$ after subtraction of the continuum, taken from Garc\'{\i}a-Barreto et al. (1996). The structures observed are: 1) two bright regions at northeast and southwest at a PA$~{\sim}166^{\circ}$, a circumnuclear emission, weak emission joining the exterior H${\alpha}$ regions and the cirumnuclear emission. Notice the lack
of H$\alpha$ emission from the compact nucleus. Contours are in arbitrary units proportional to the peak.

FIGURE 3

Superposition of H${\alpha}$, in grey scale, upon continuum 
emission, in contours. Notice that the SE bright source lies at the position where the optical continuum isophotes show a relative lack of emission compared to the surrounding regions (see also Figure 1b). The position angle of the 
line joining the two H$\alpha$ sources straddling the nucleus is P.A.$\simeq165^{\circ}$ and differs by about $\simeq15^{\circ}$ from the P.A. of the elongated isophotes of the stellar bar.

FIGURE 4

Schematic diagram showing the components of the mass distribution. The $x$ axis is along the major axis of the bar of the galaxy. The plane of the sky is represented by the plane formed by {\bf ($x', y'$)}. The $x'$ axis is along the projected major axis of the disk of the galaxy. The line of sight from the observer to the galaxy is along +$z'$.

FIGURE 5

a) Optical red ({\it I}) continuum brightness distribution observed along the axis $x'=0$. Scale units are arbitrary but are proportional to the true brightness by a constant factor. b) Brightness distribution obtained from our models along the axis $x'=0$. The continuous line is the output from our model with major semi-axes $a_{{\rm BRI}}=1$ kpc and $a_{{\rm BRII}}=1.3$ kpc, while the dotted line is the output with  $a_{{\rm BRI}}=1$ kpc and $a_{{\rm BRII}}=1.5$ kpc. In both models the density of the second component of the bar is $\rho_{{\rm BRII}}(a)~=~q_{{\rm BRII}}+p_{{\rm BRII}}a^4$. Similarly, (c)-(d), (e)-(f), (g)-(h), (i)-(j), and (k)-(l) are the observed and model brightnesses along $x'=0.2, 0.4, 0.6, 0.8, 1.2$ and 1.4 kpc, respectively.

FIGURE 6 

Brightnesses as in Fig. 5 but now along the lines $y'= 0, 0.2, 0.4,$ and 0.8 kpc.

FIGURE 7

a) Isophotes in the first fit ($a_{{\rm BRI}}=1$ kpc, $a_{{\rm BRII}}=1.3$ kpc) using the three forms of the mass density in the second component of the bar (boxy isophotes). The outer curve in a given isophote is the output of the model using the third form of the density in Eq. (3); the corresponding inner curve is the output of the model using the first form of the density in Eq. (3). b) Isophotes in the second fit ($a_{{\rm BRI}}=1$ kpc, $a_{{\rm BRII}}=1.5$ kpc) using the third form of the mass density in the second component of the bar (boxy isophotes). c) Isophotes in the second fit with $\phi=90^{\circ}$ and  $\theta=90^{\circ}$, as if the galaxy were viewed edge on. An axial ratio $\xi_D=0.05$ is considered in the disk to see more clearly the boxy structure. d) A close up of the central region in c). Notice that in all four figures the $x'$ axis is along the projected major axis of the disk of the galaxy.

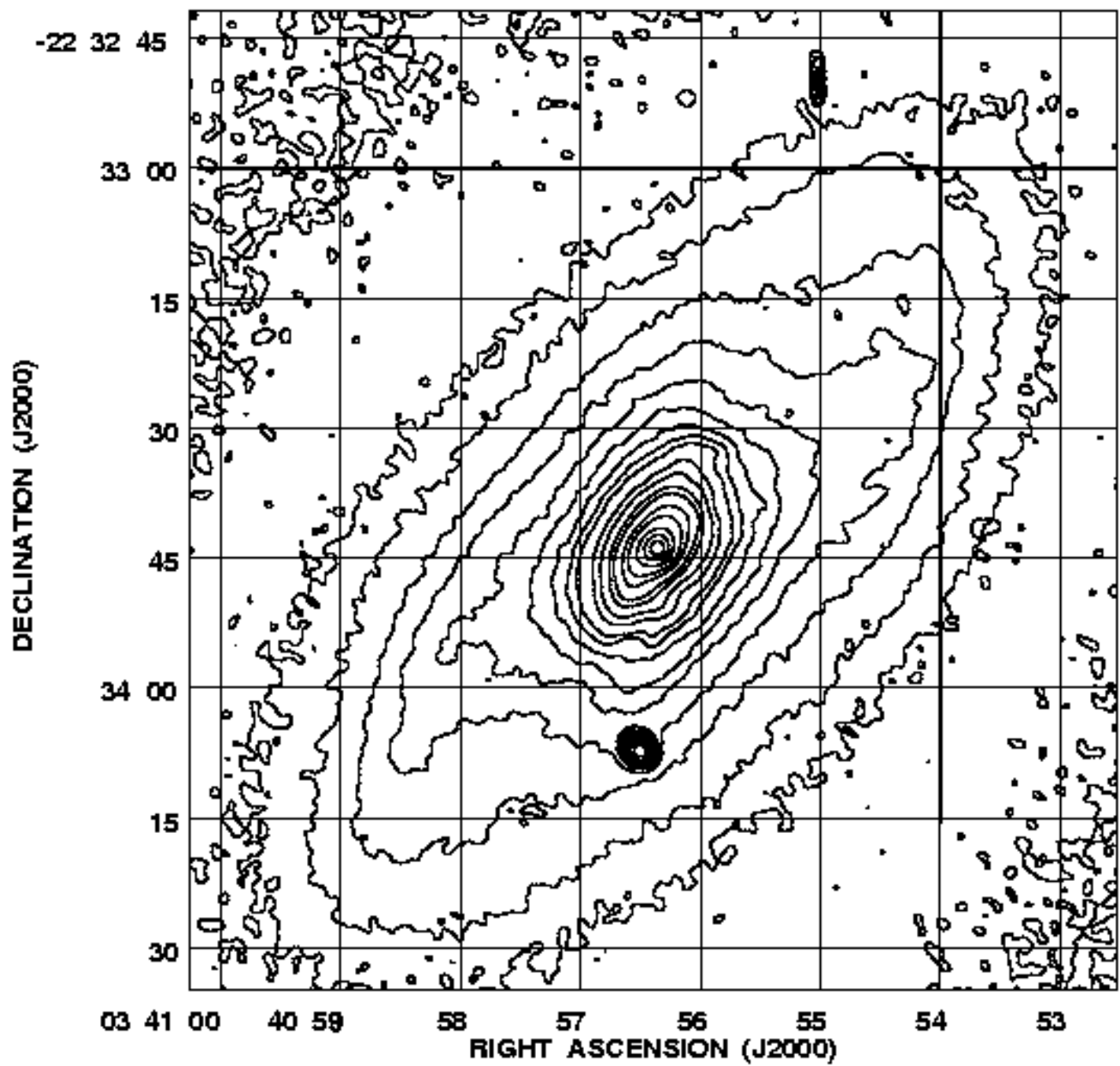
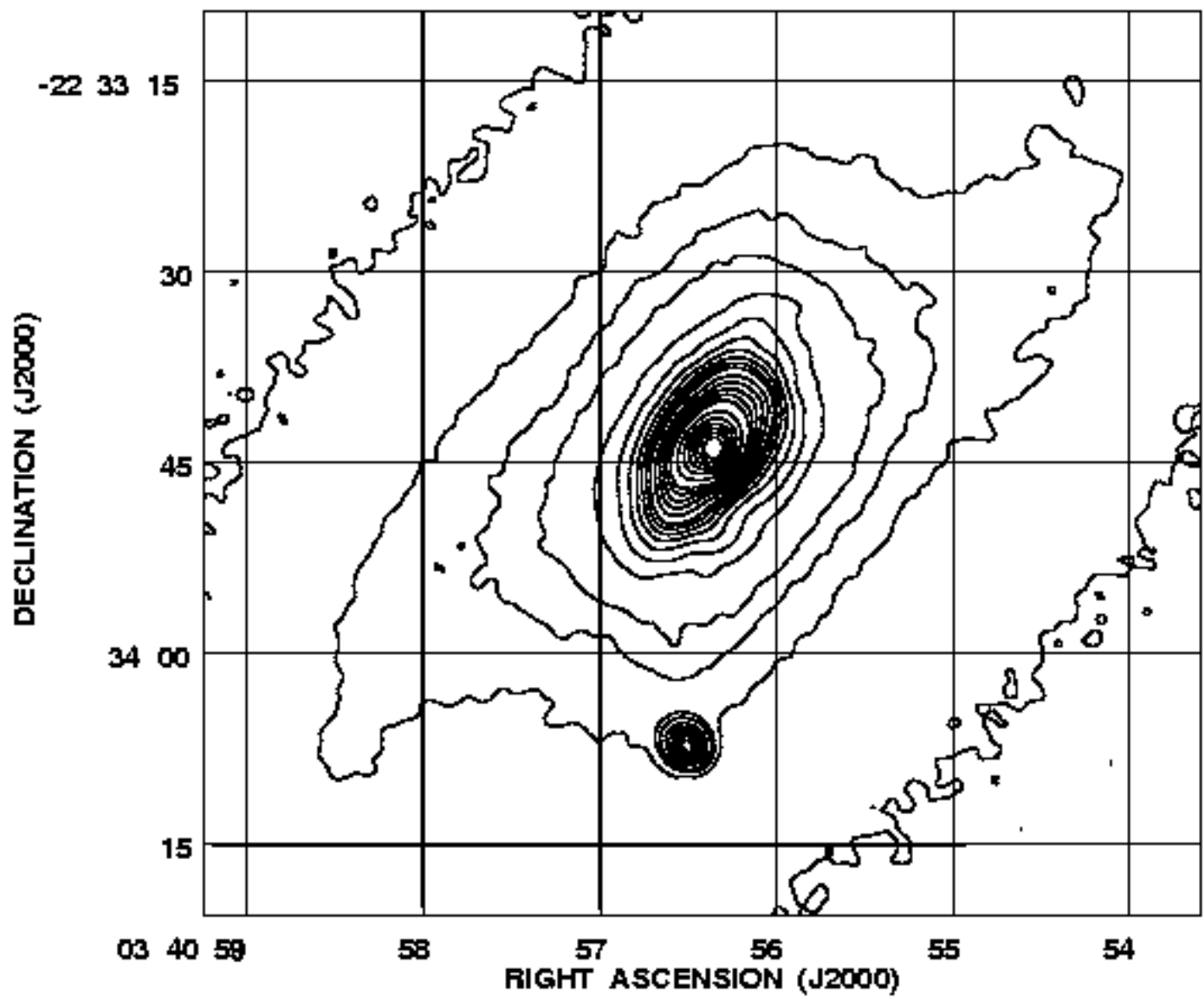
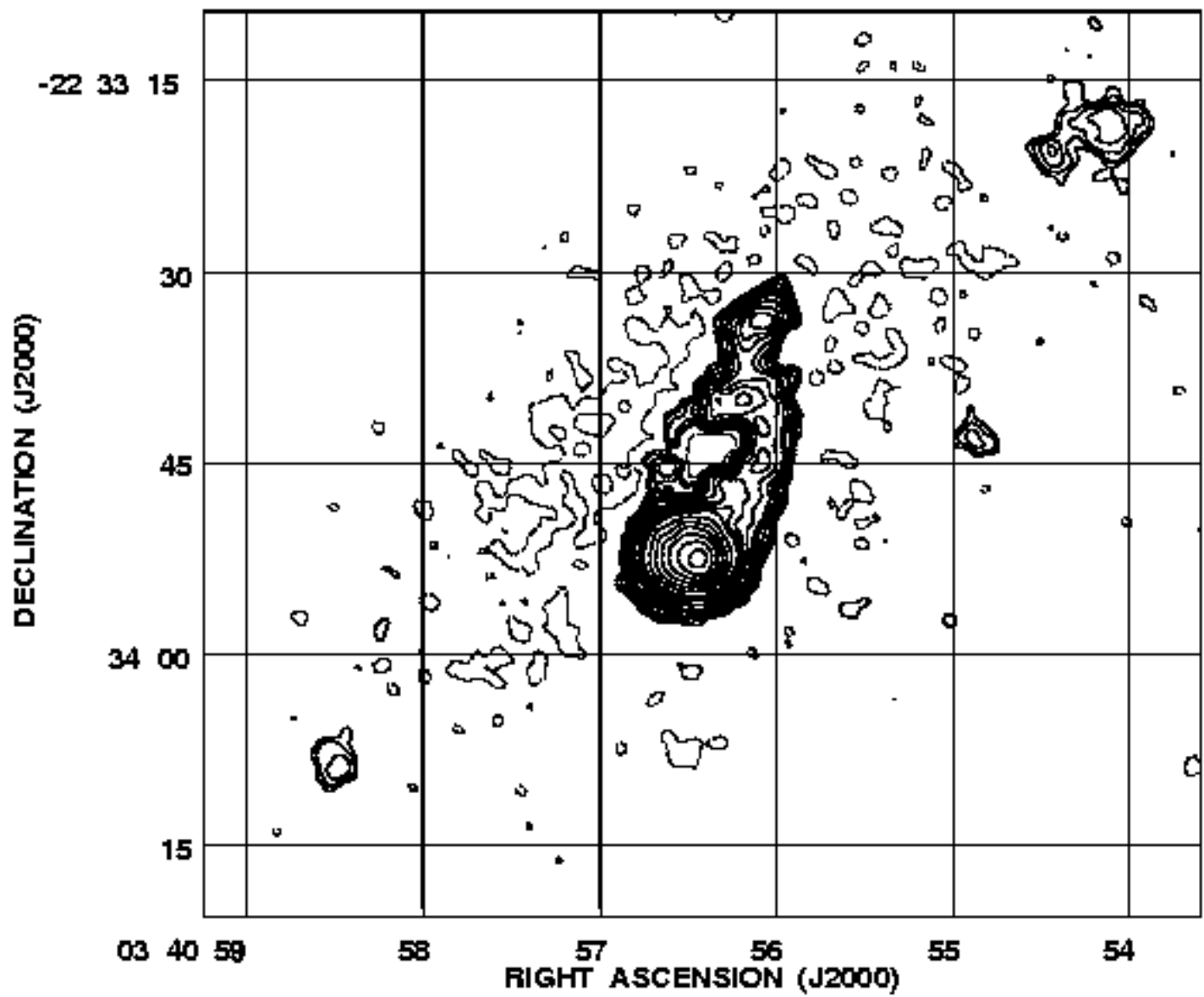
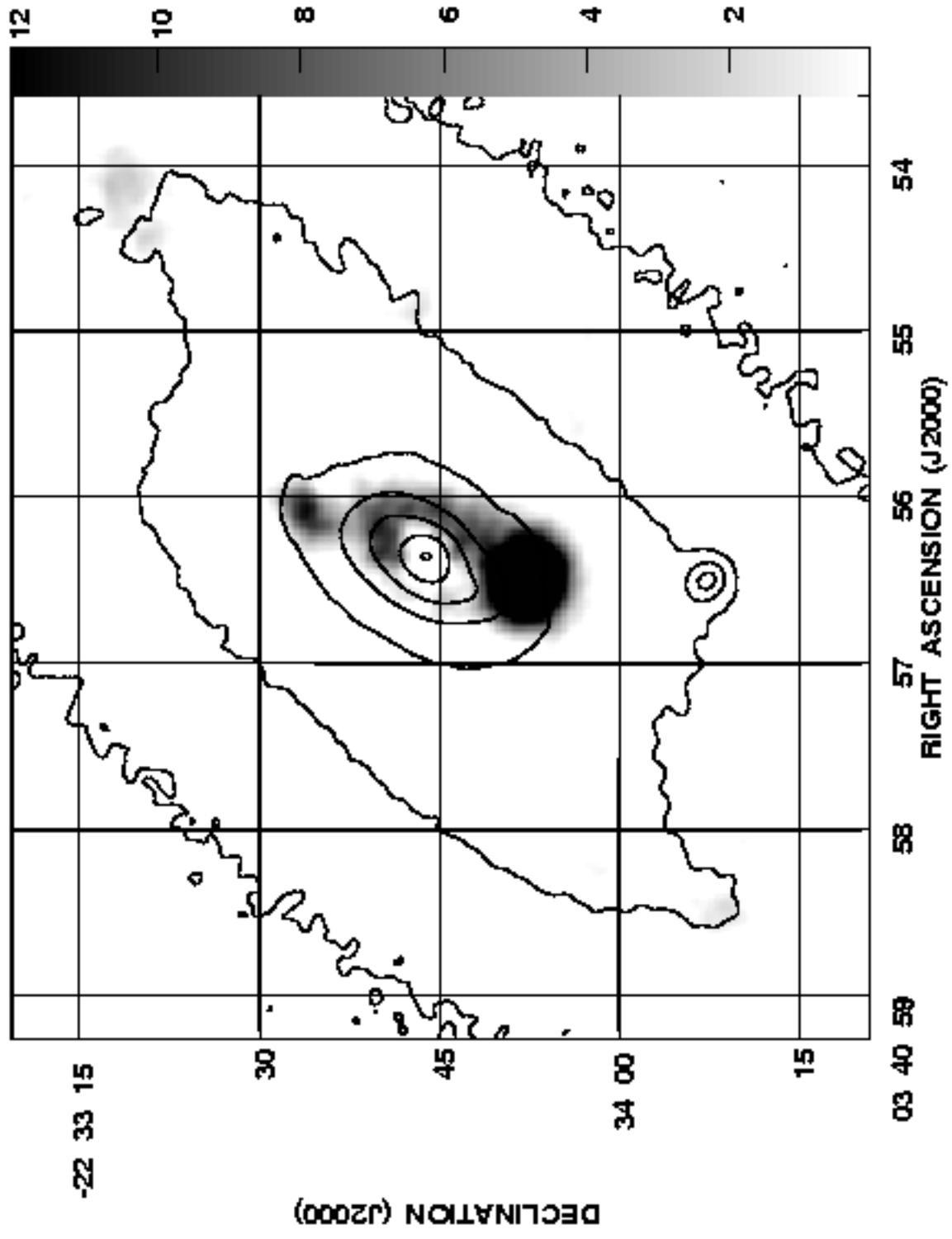
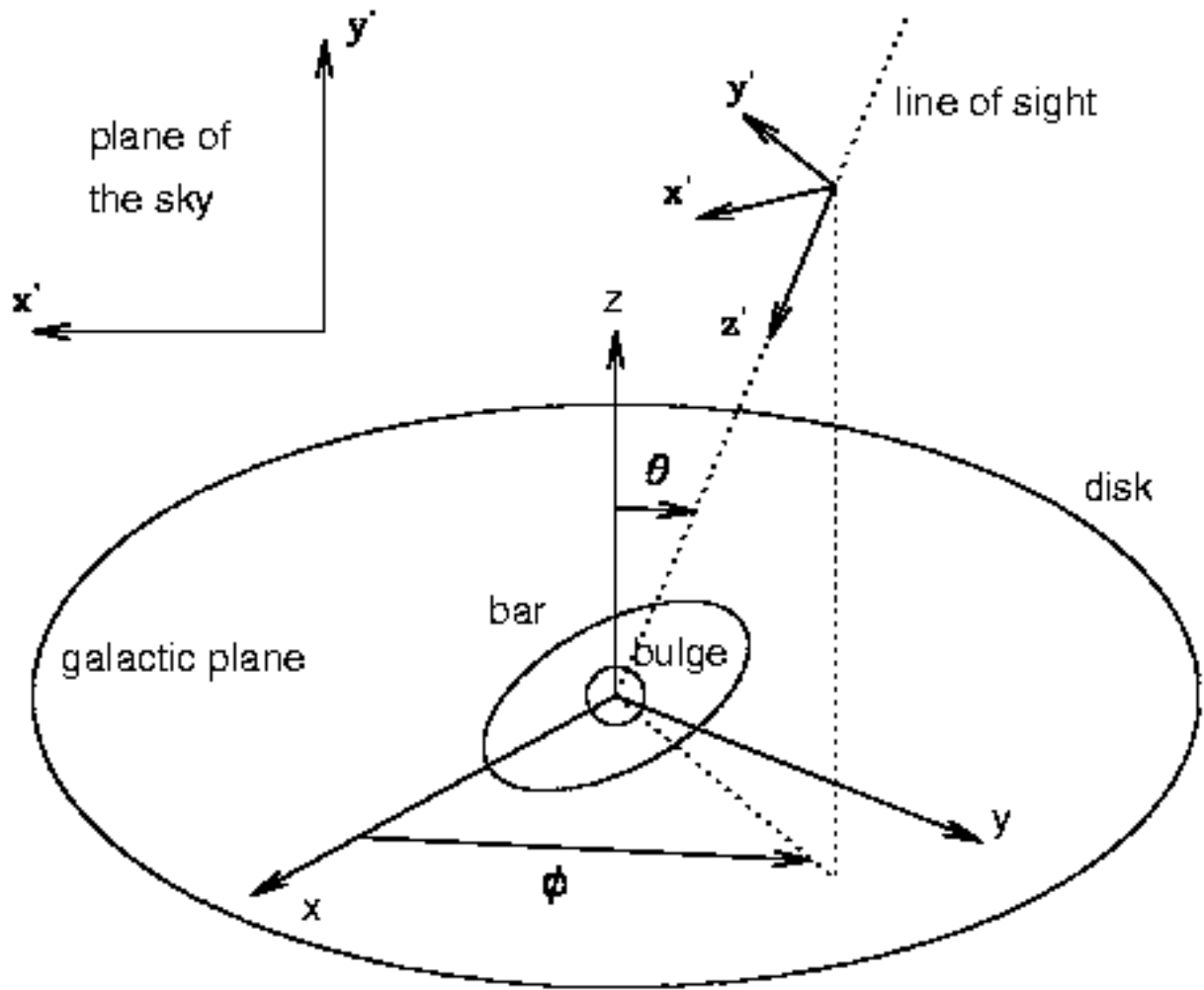
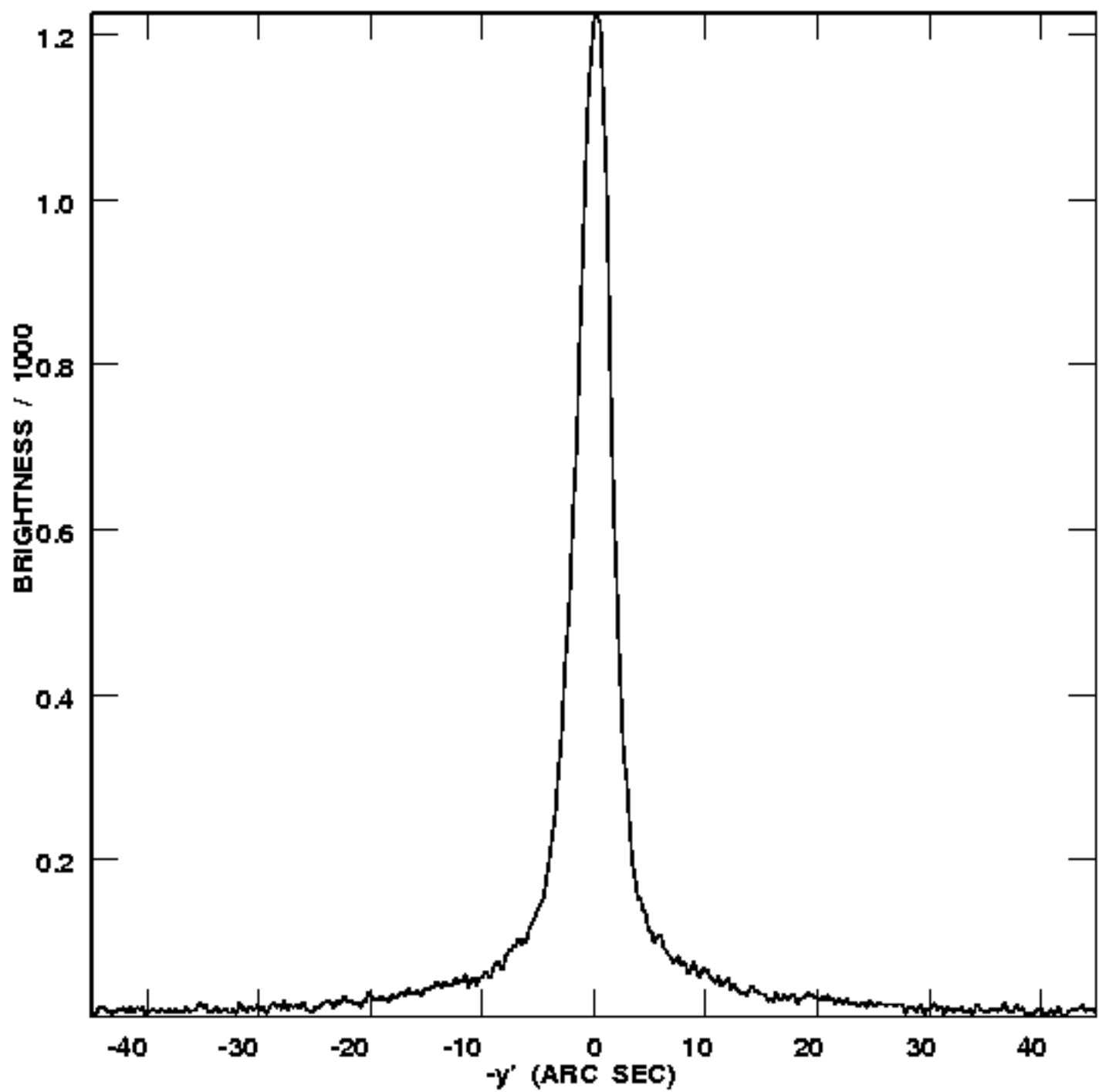
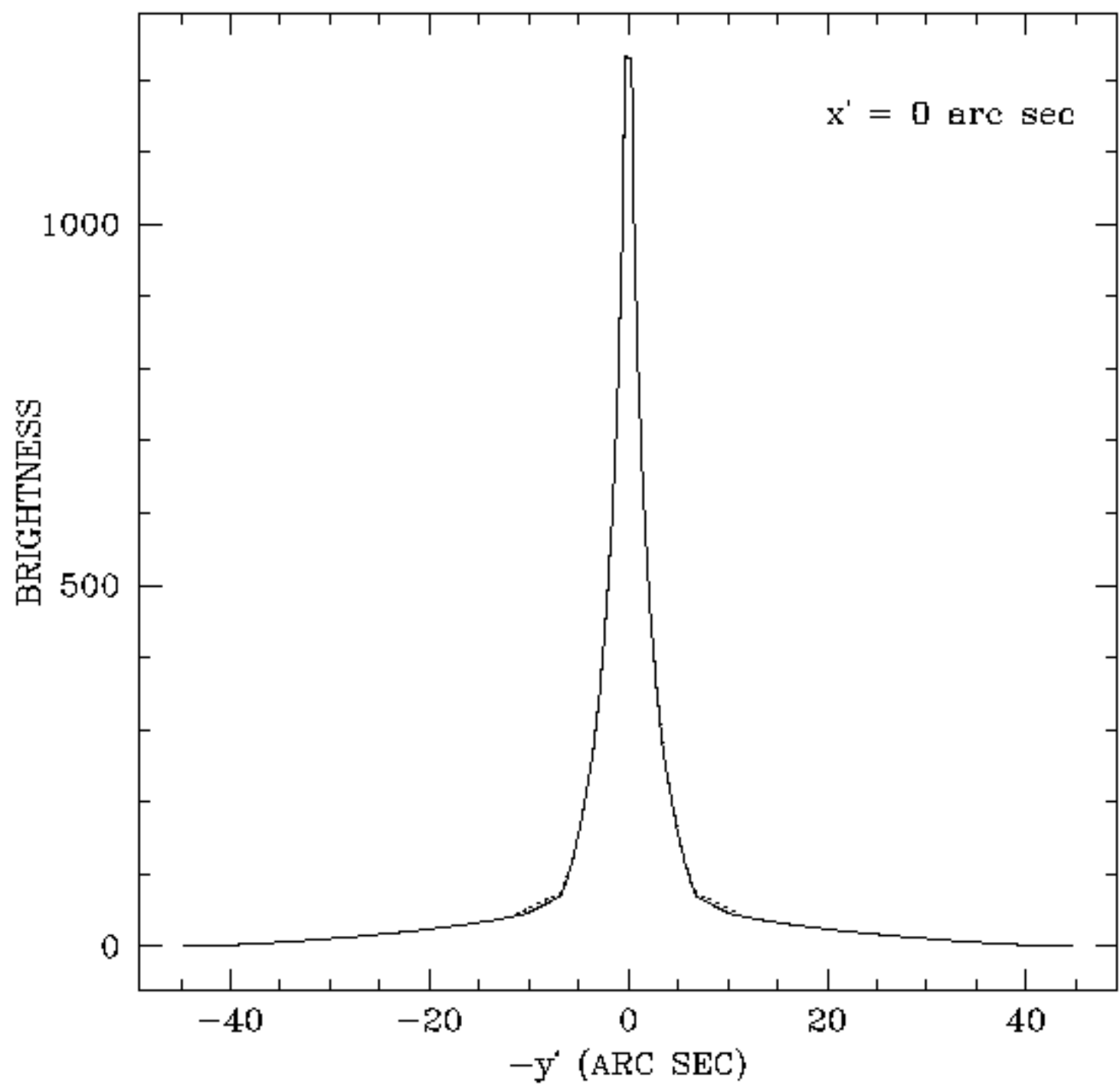
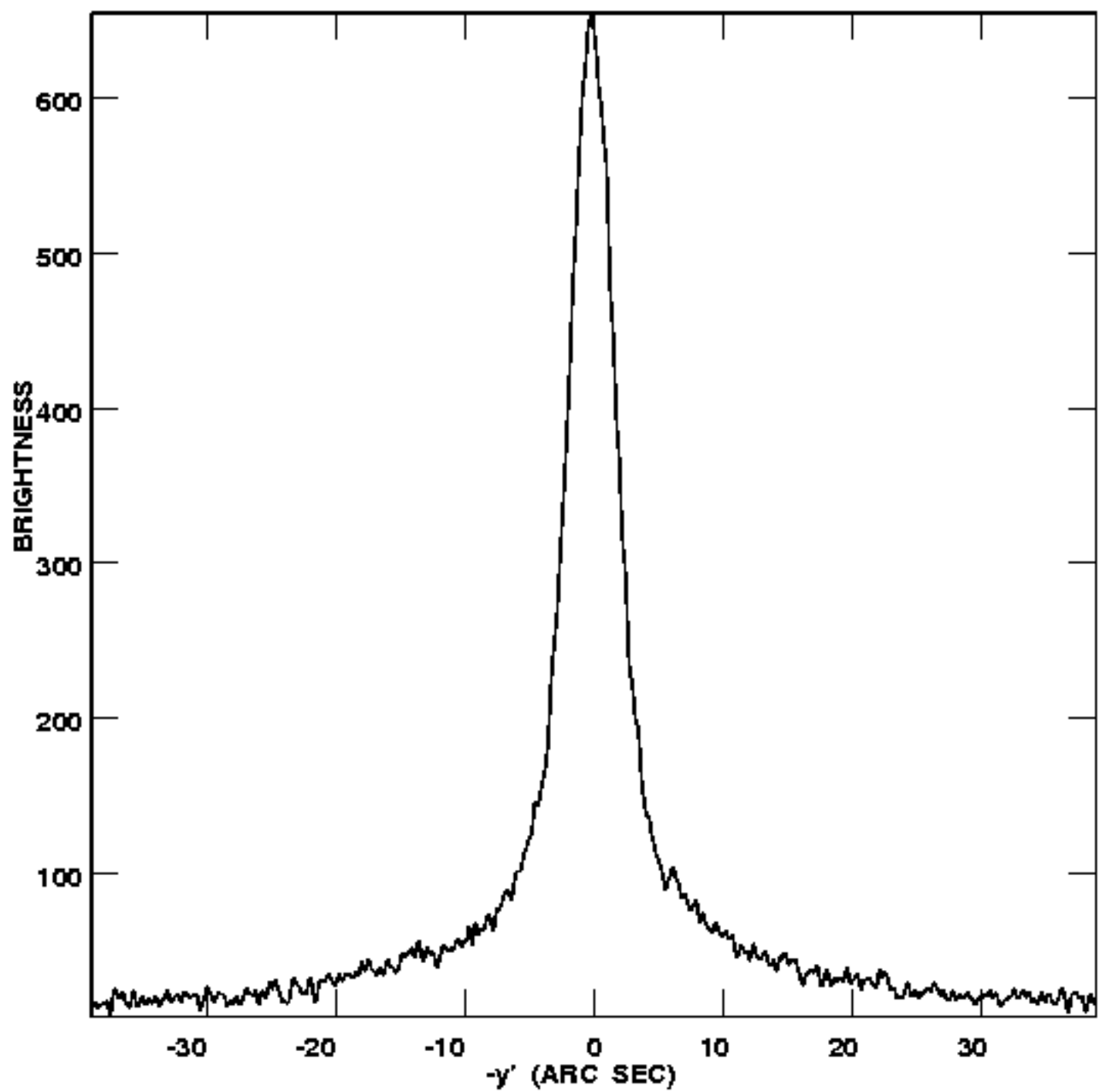
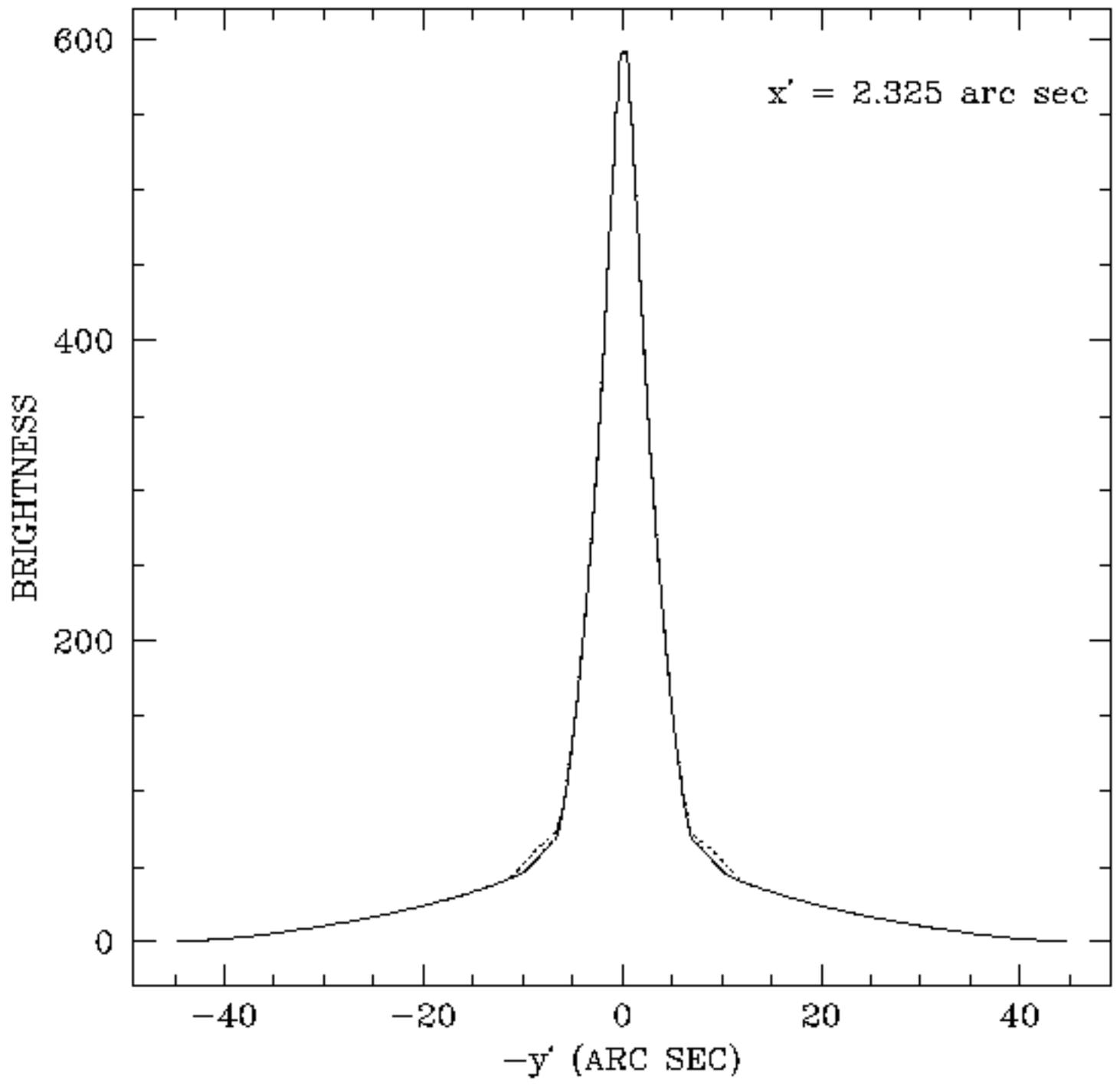
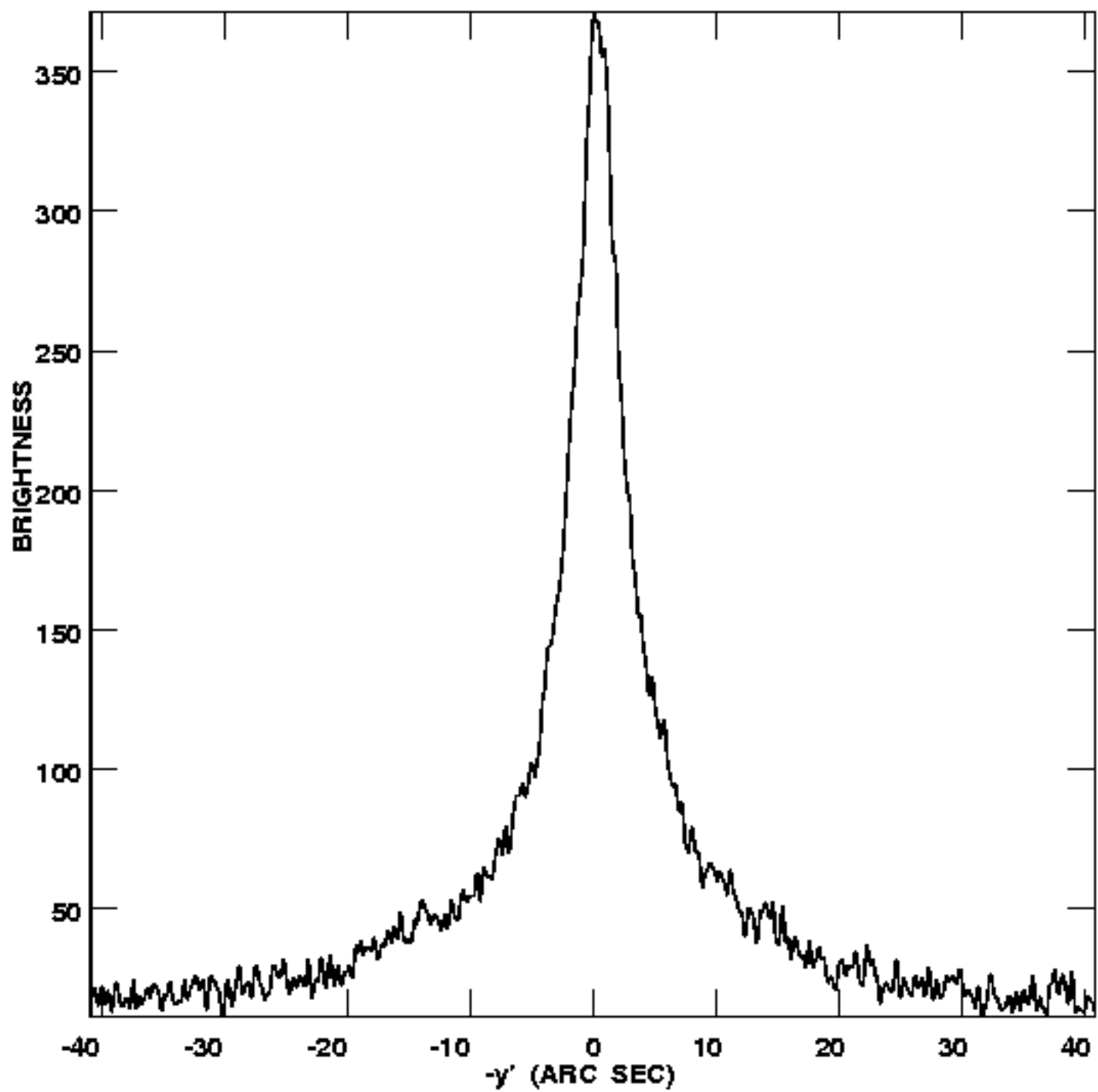
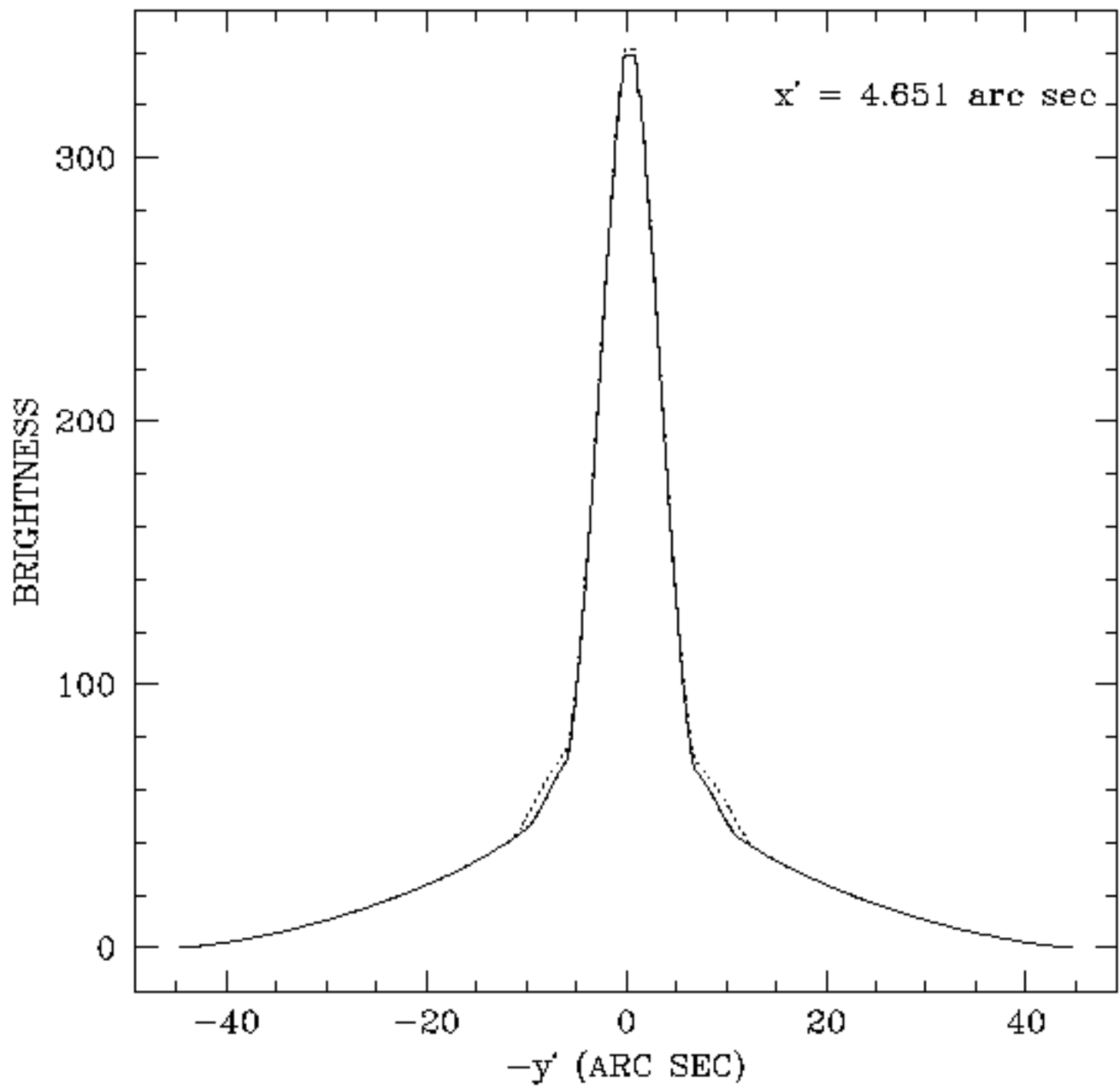
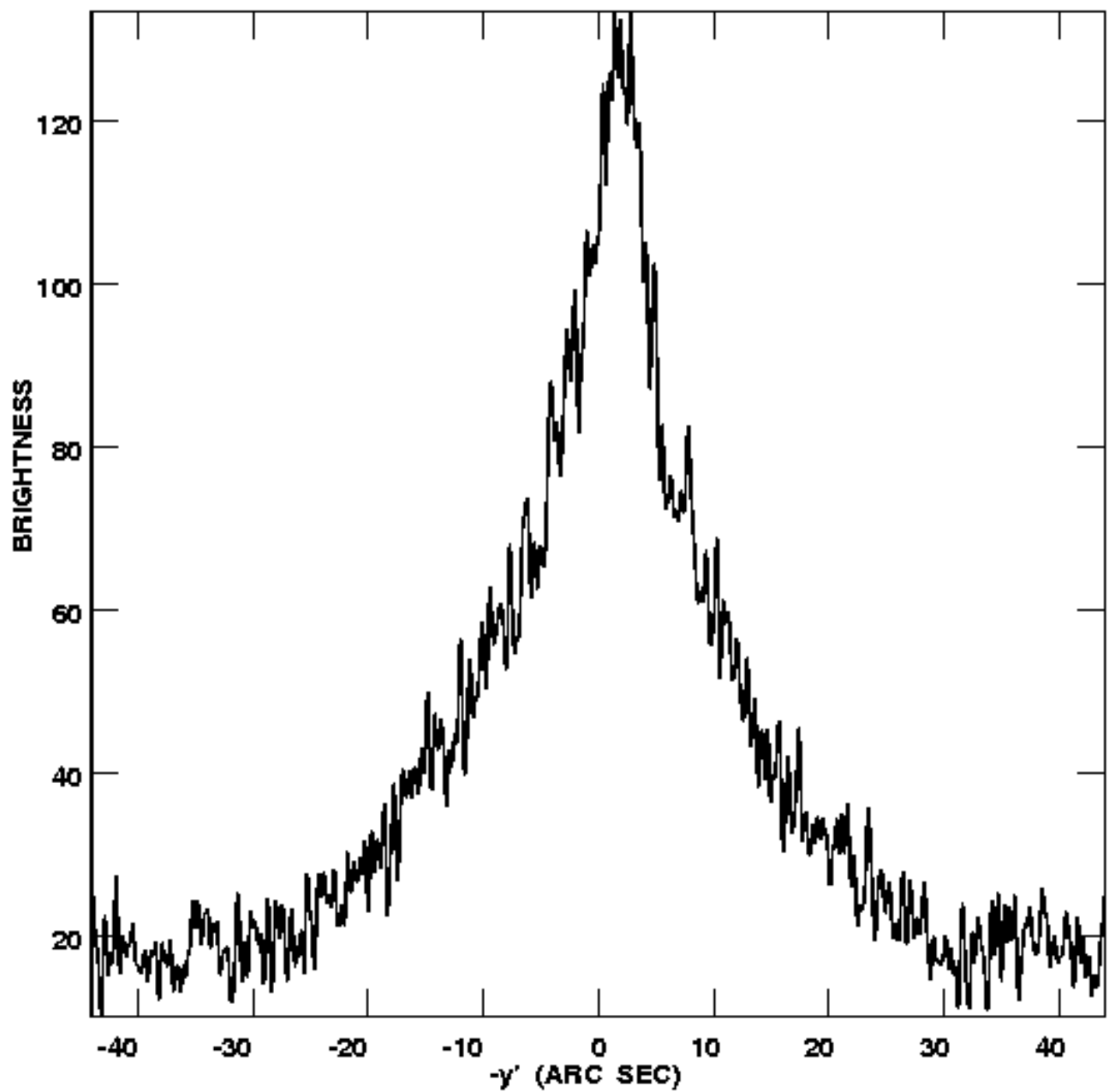
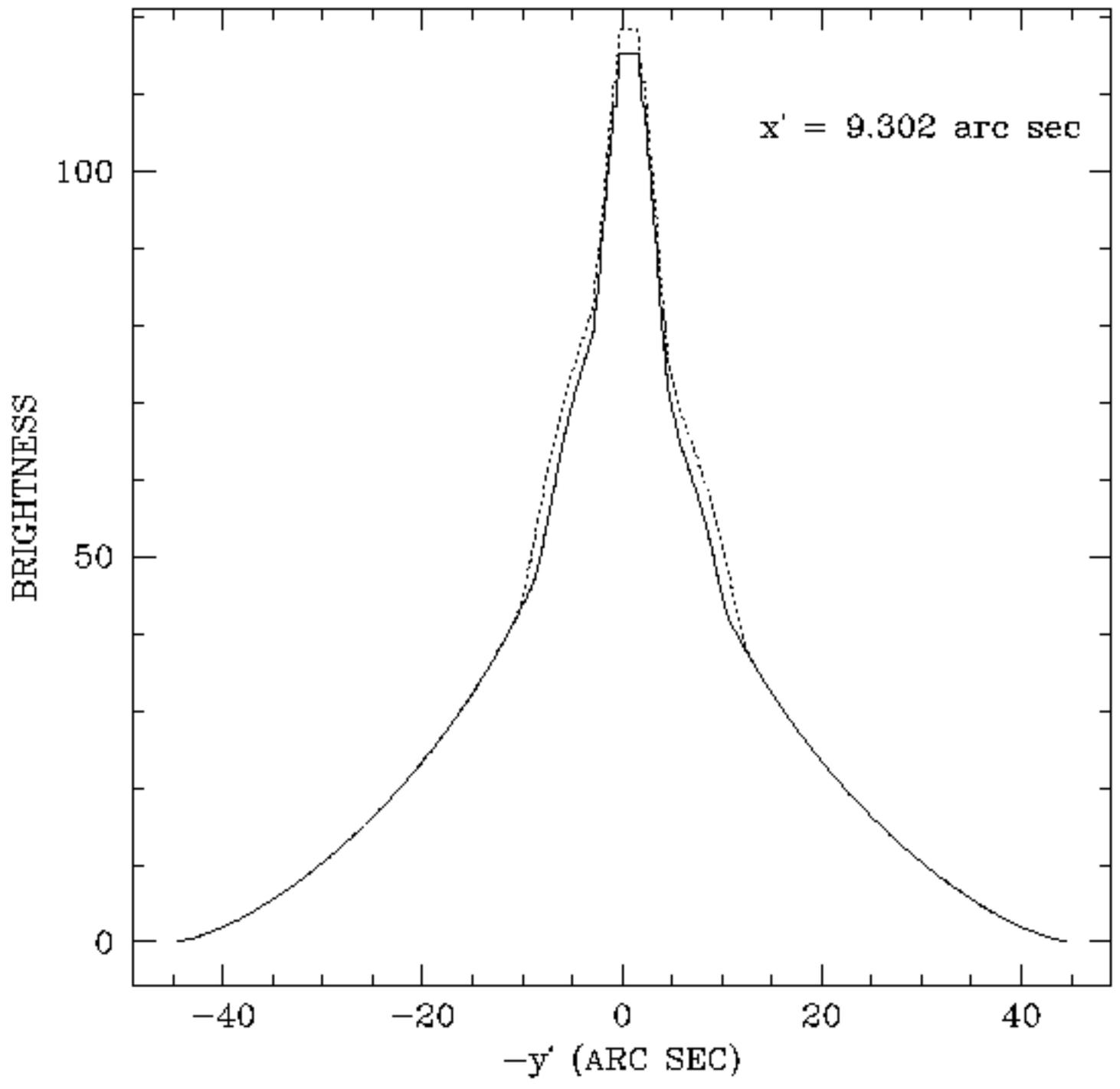
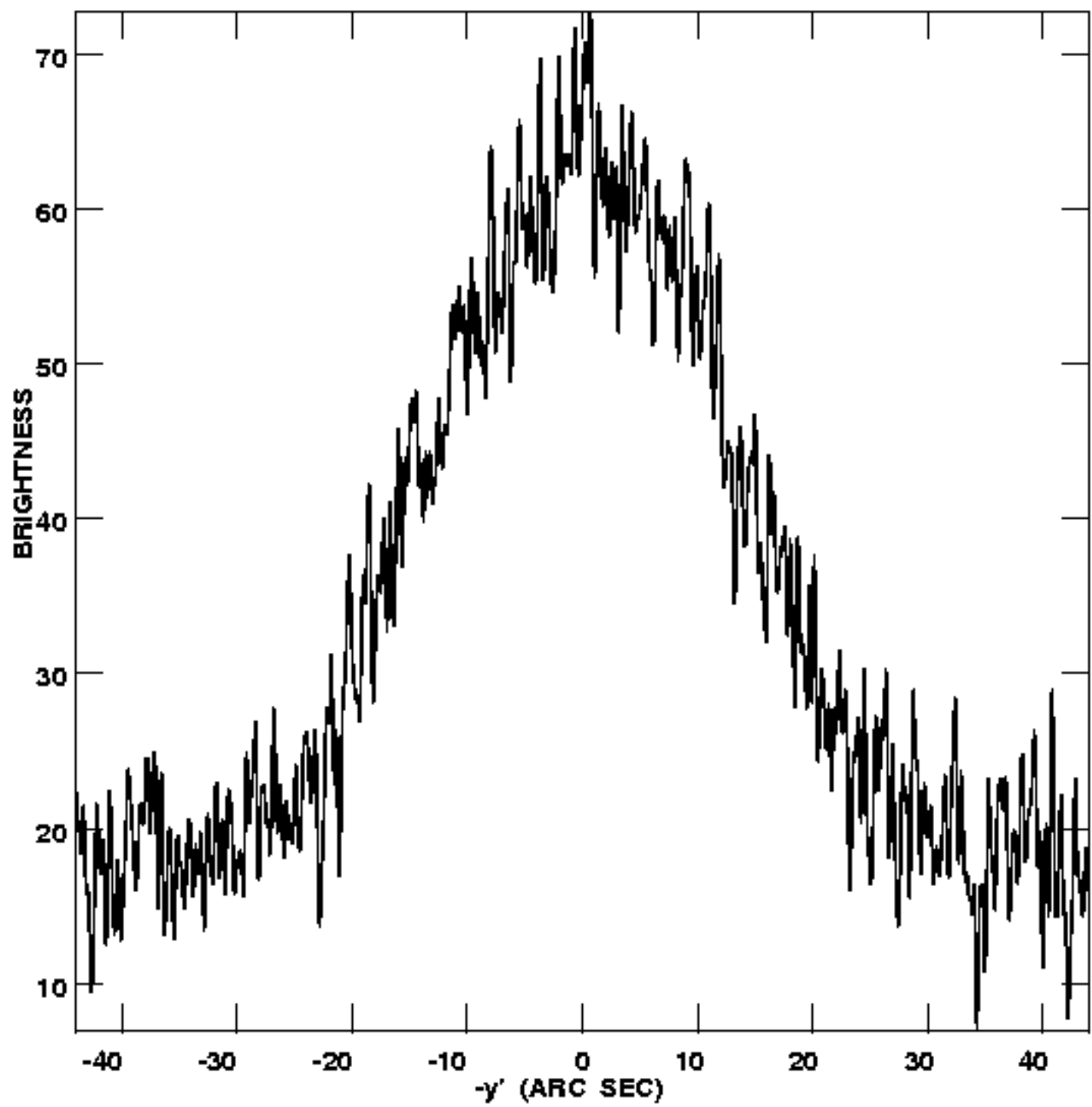
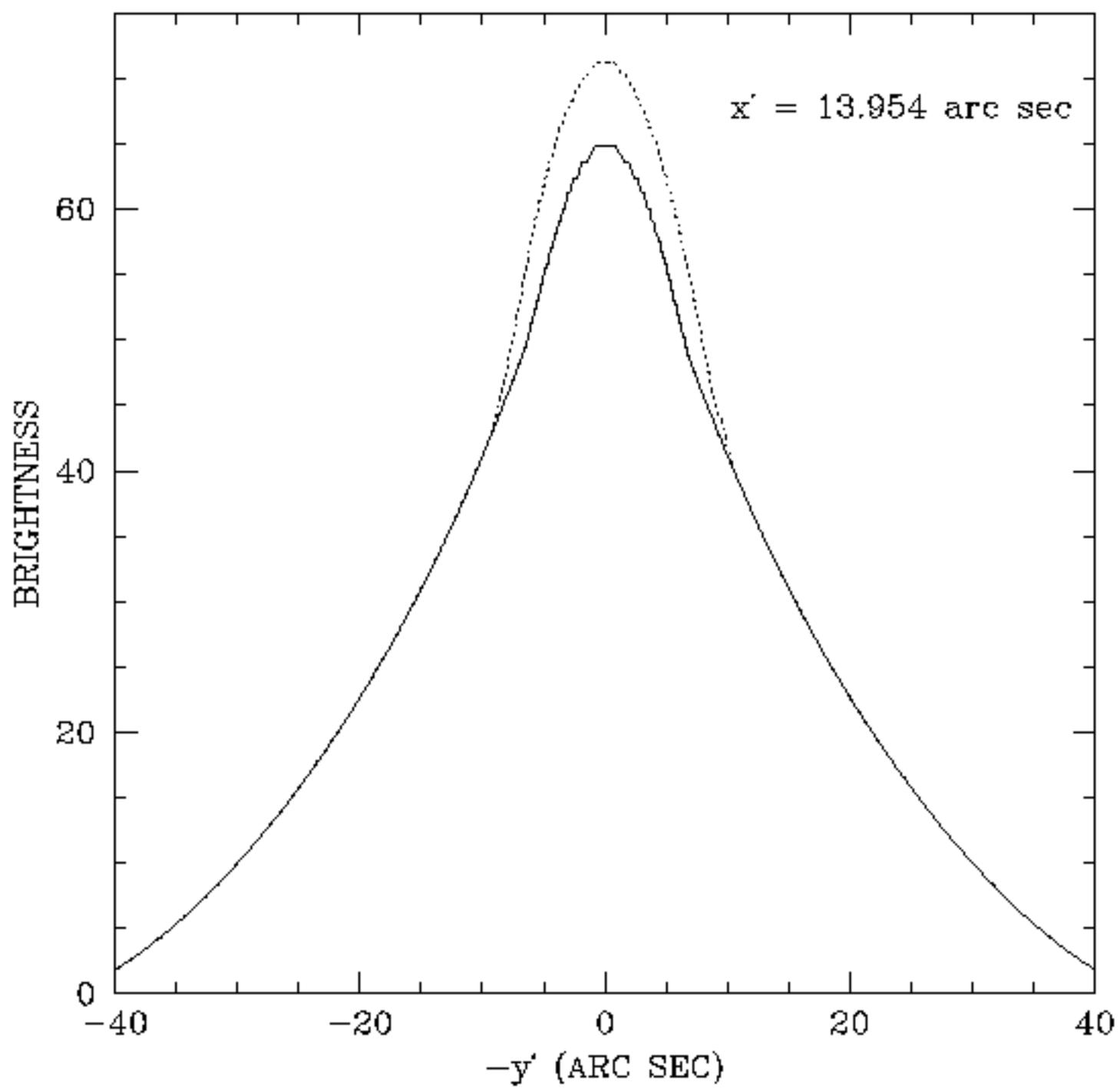
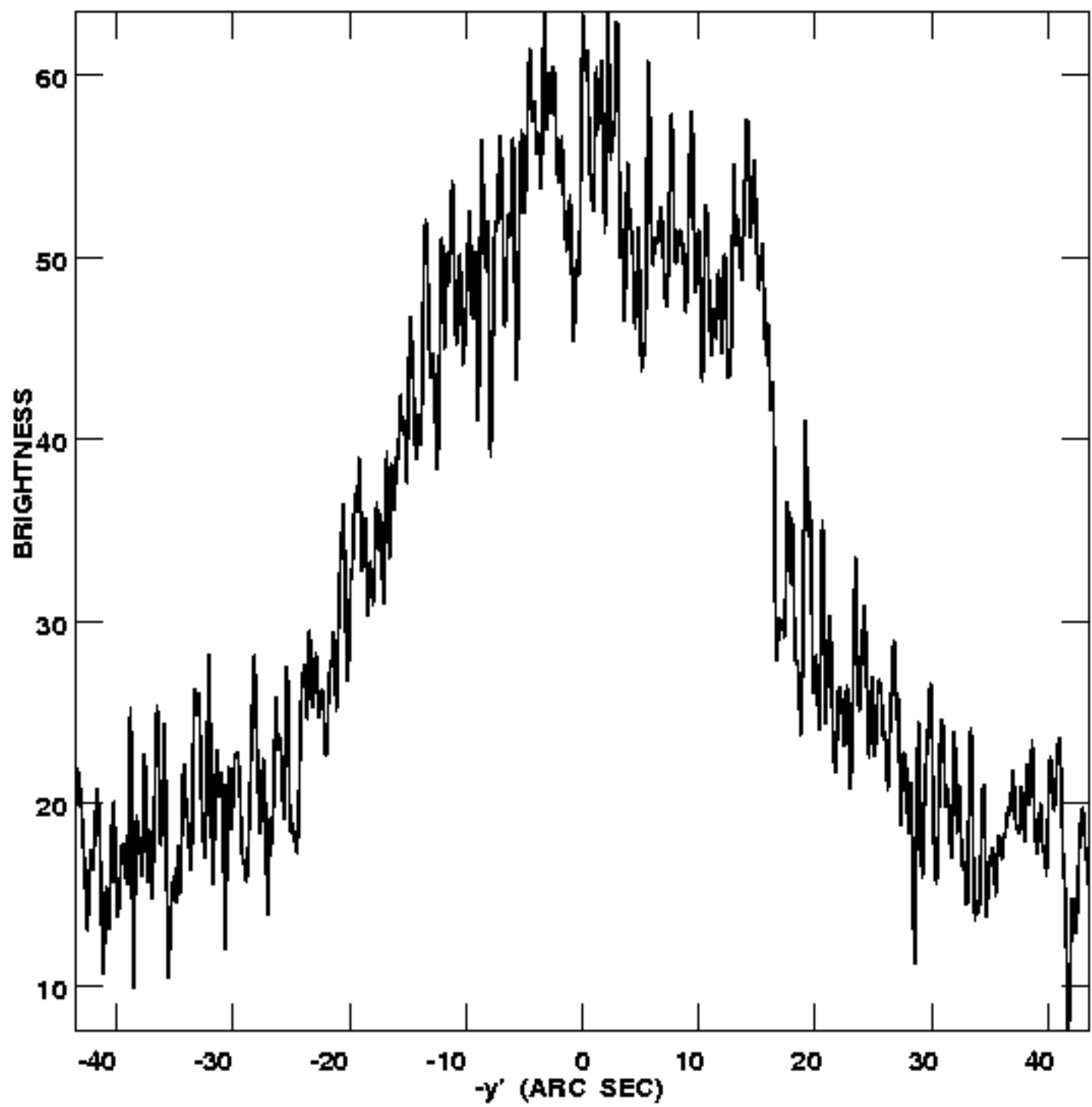
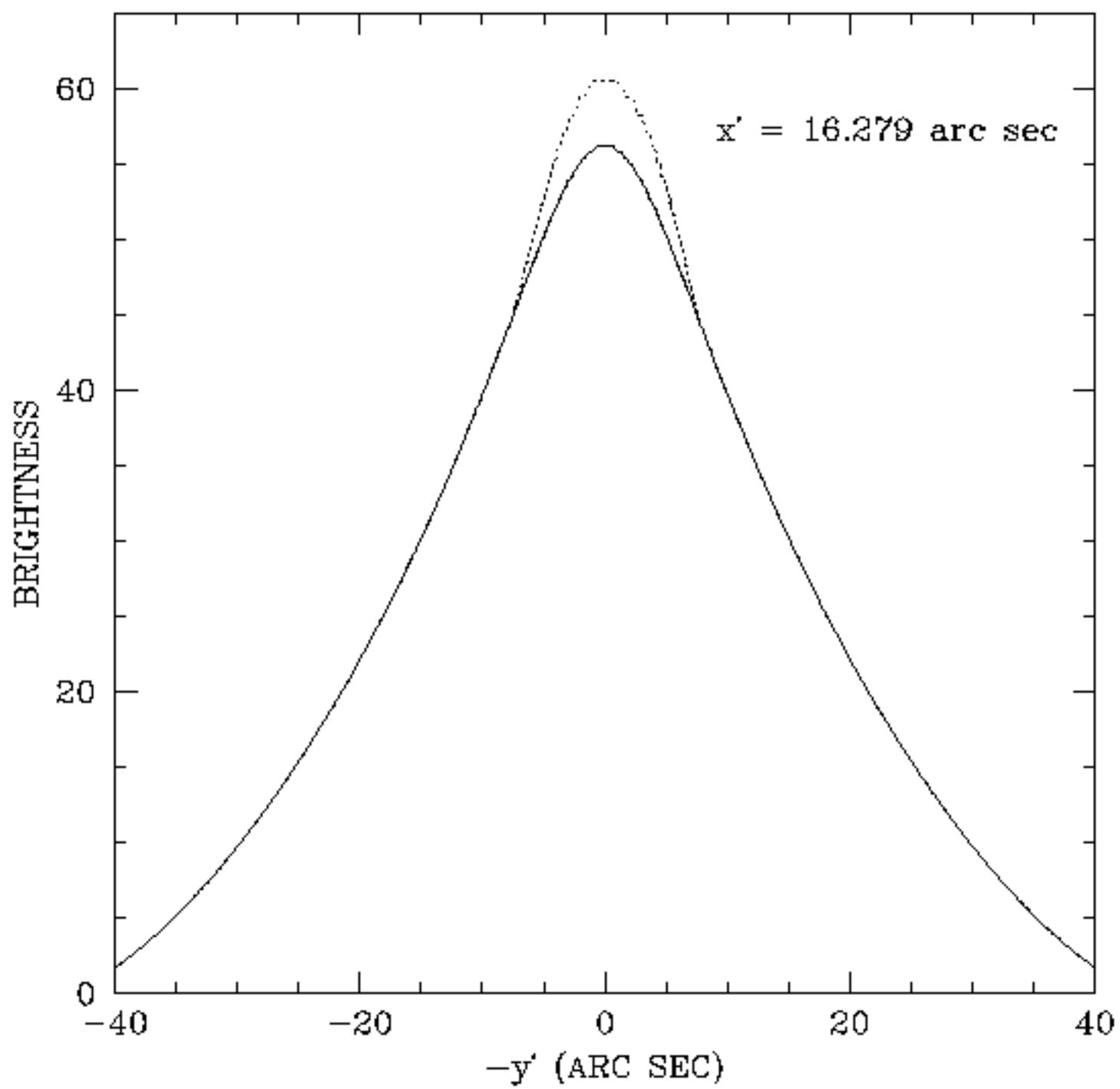
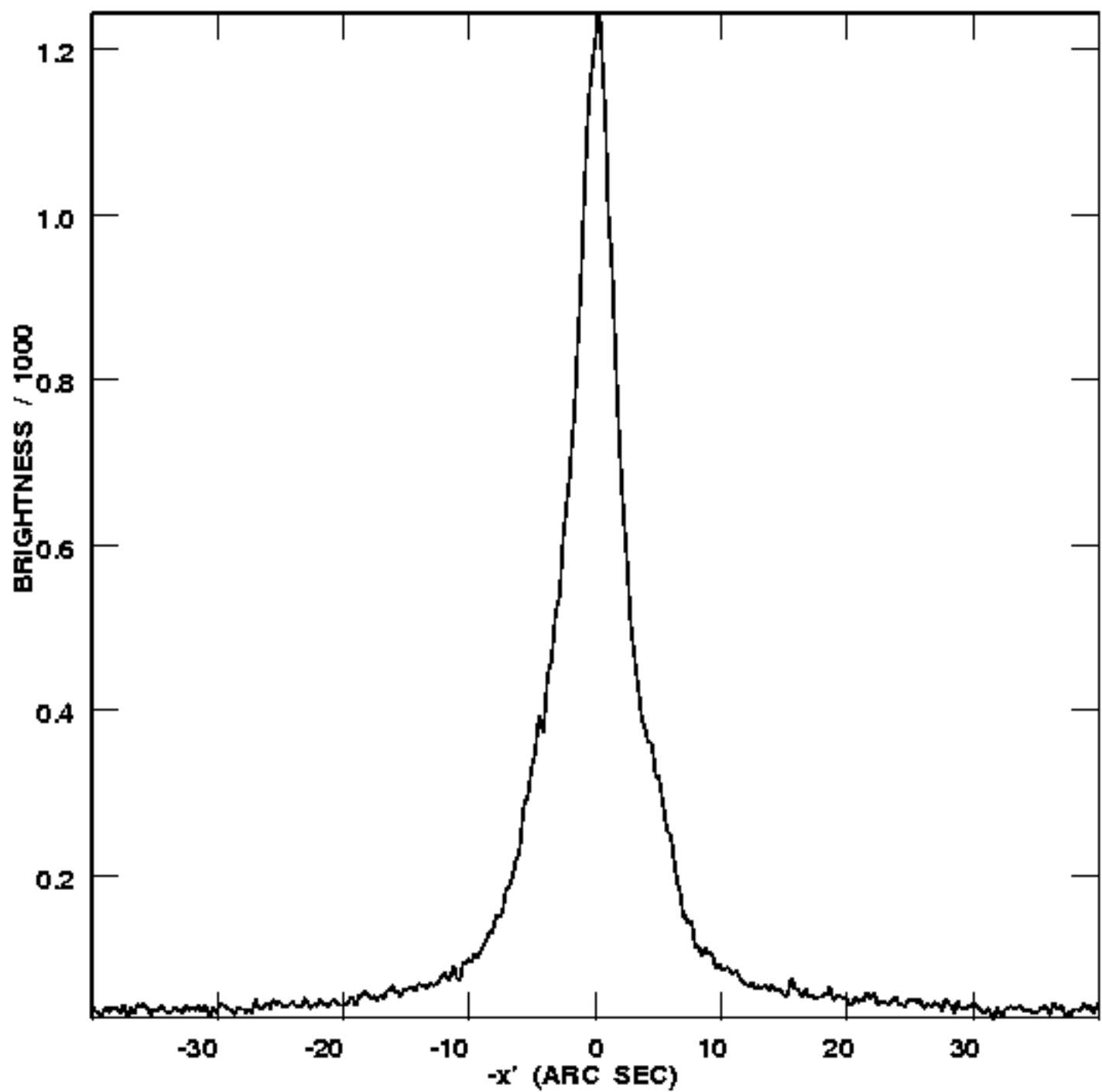
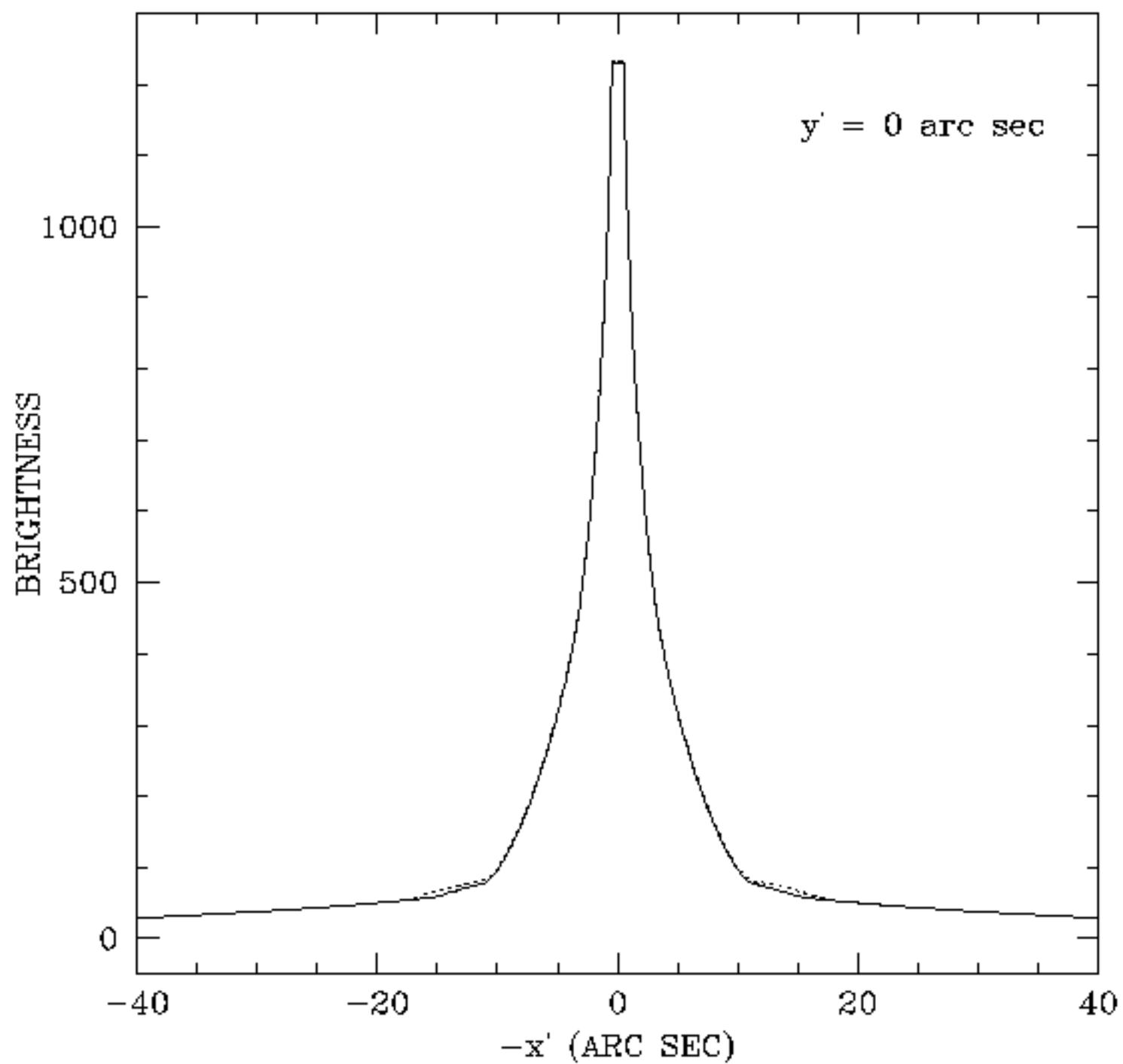
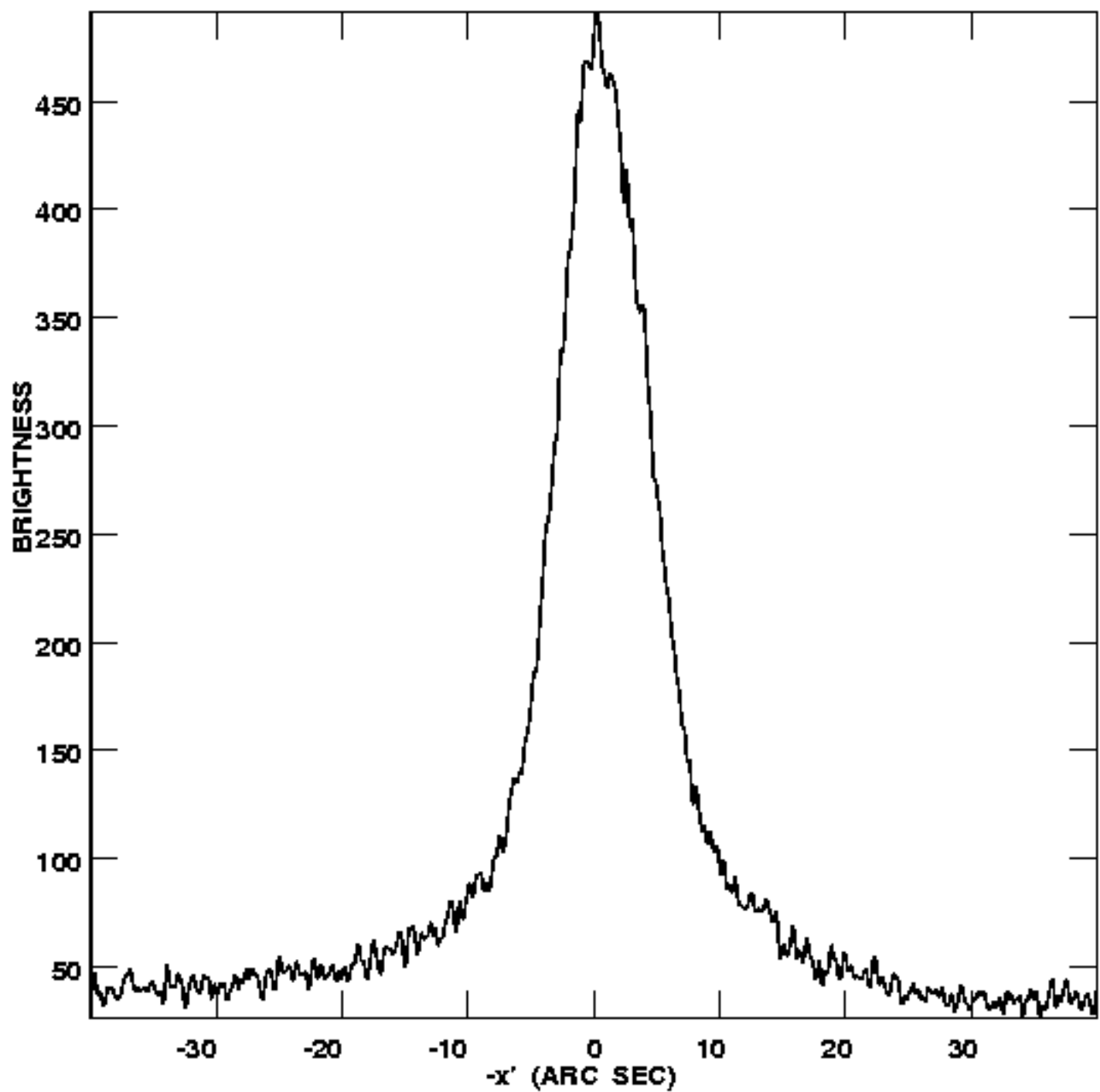
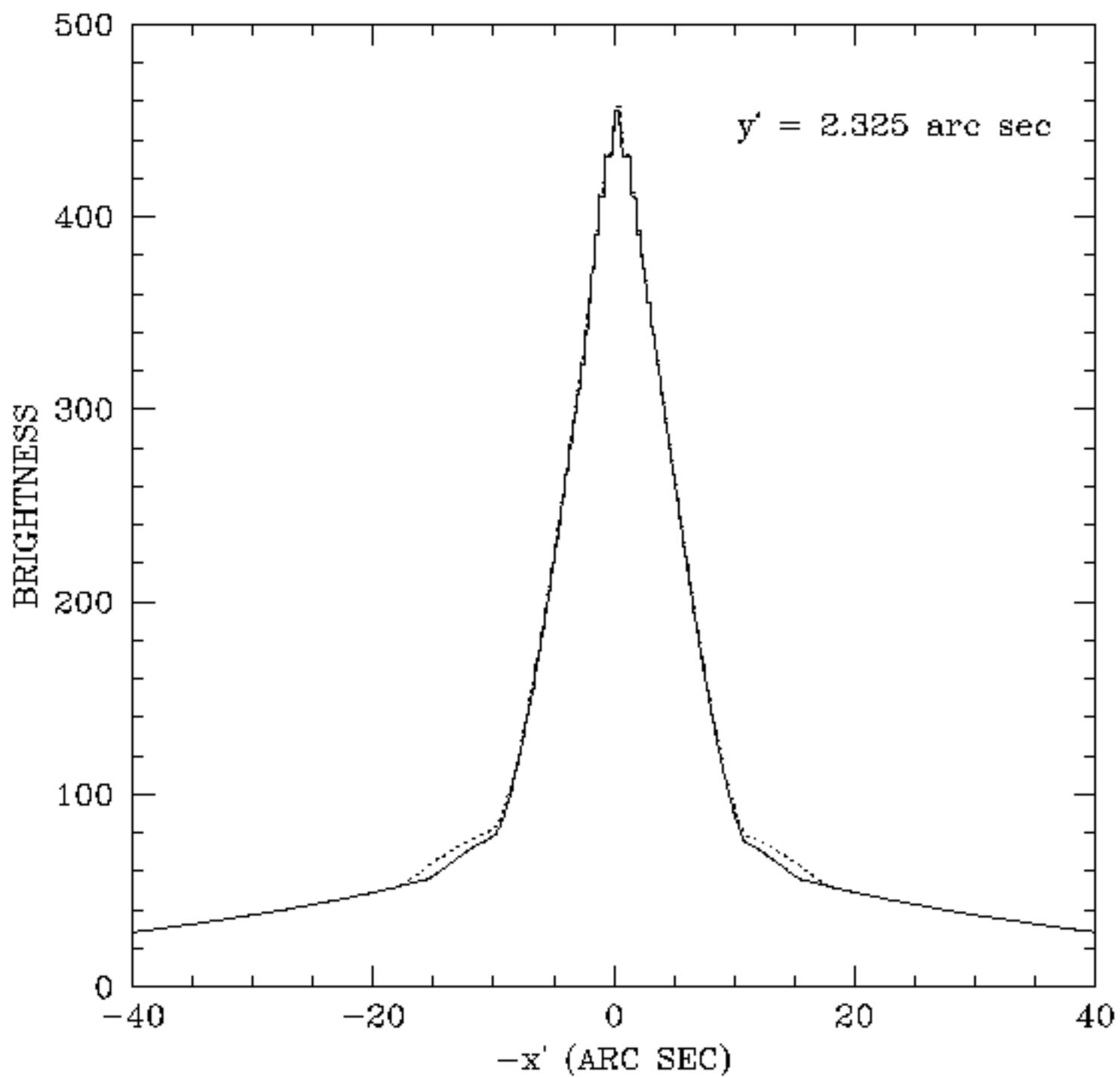
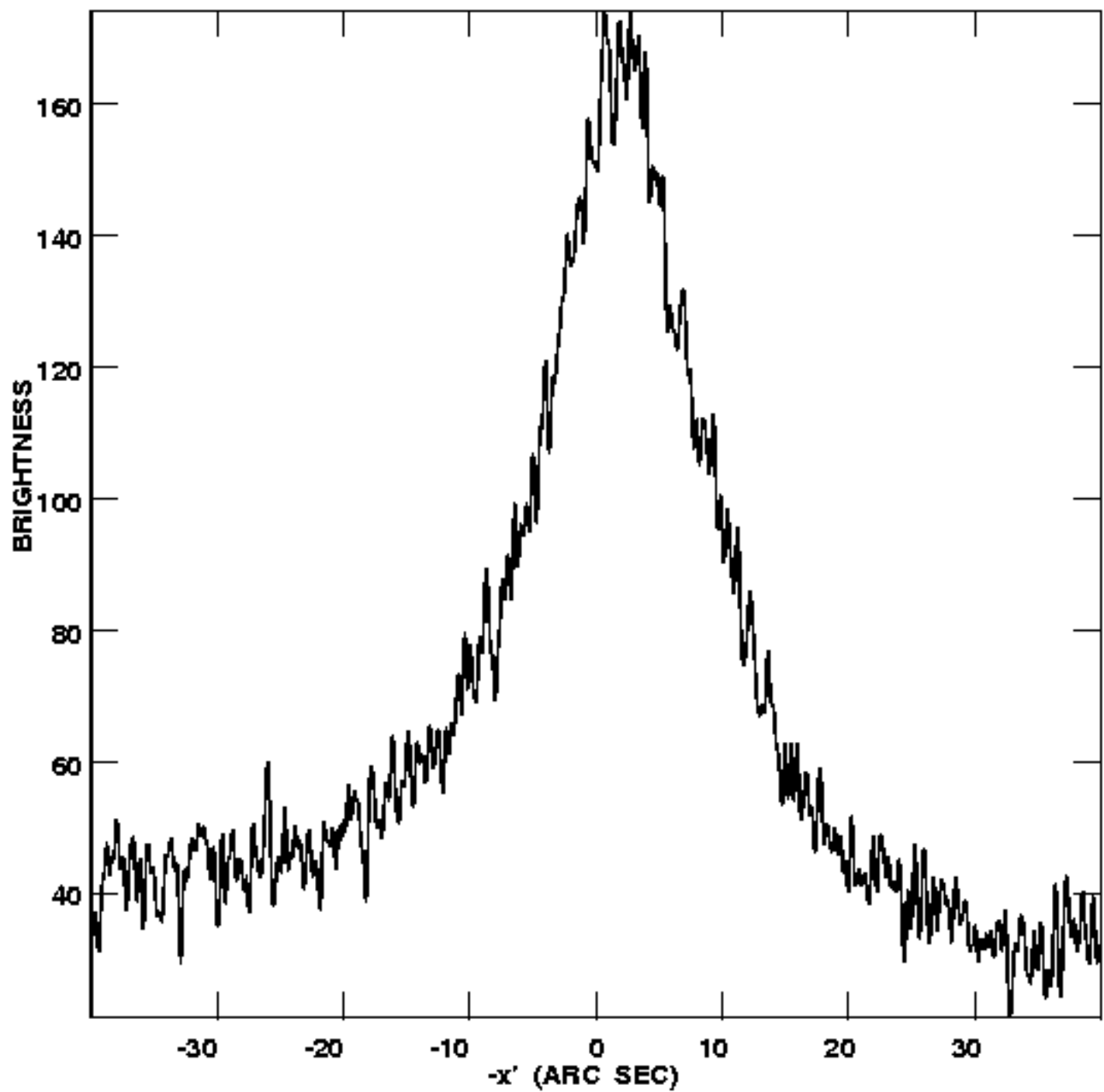
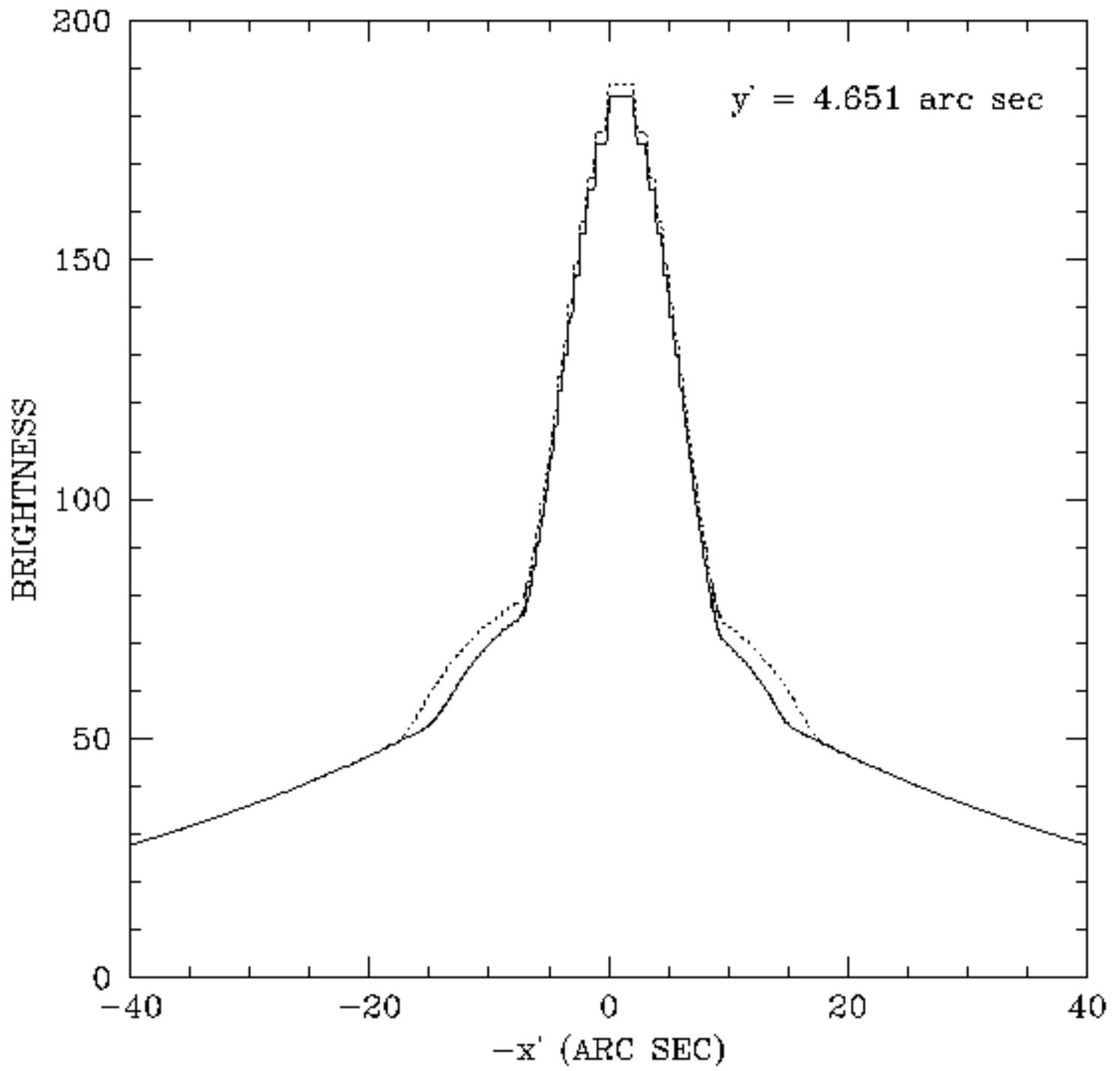
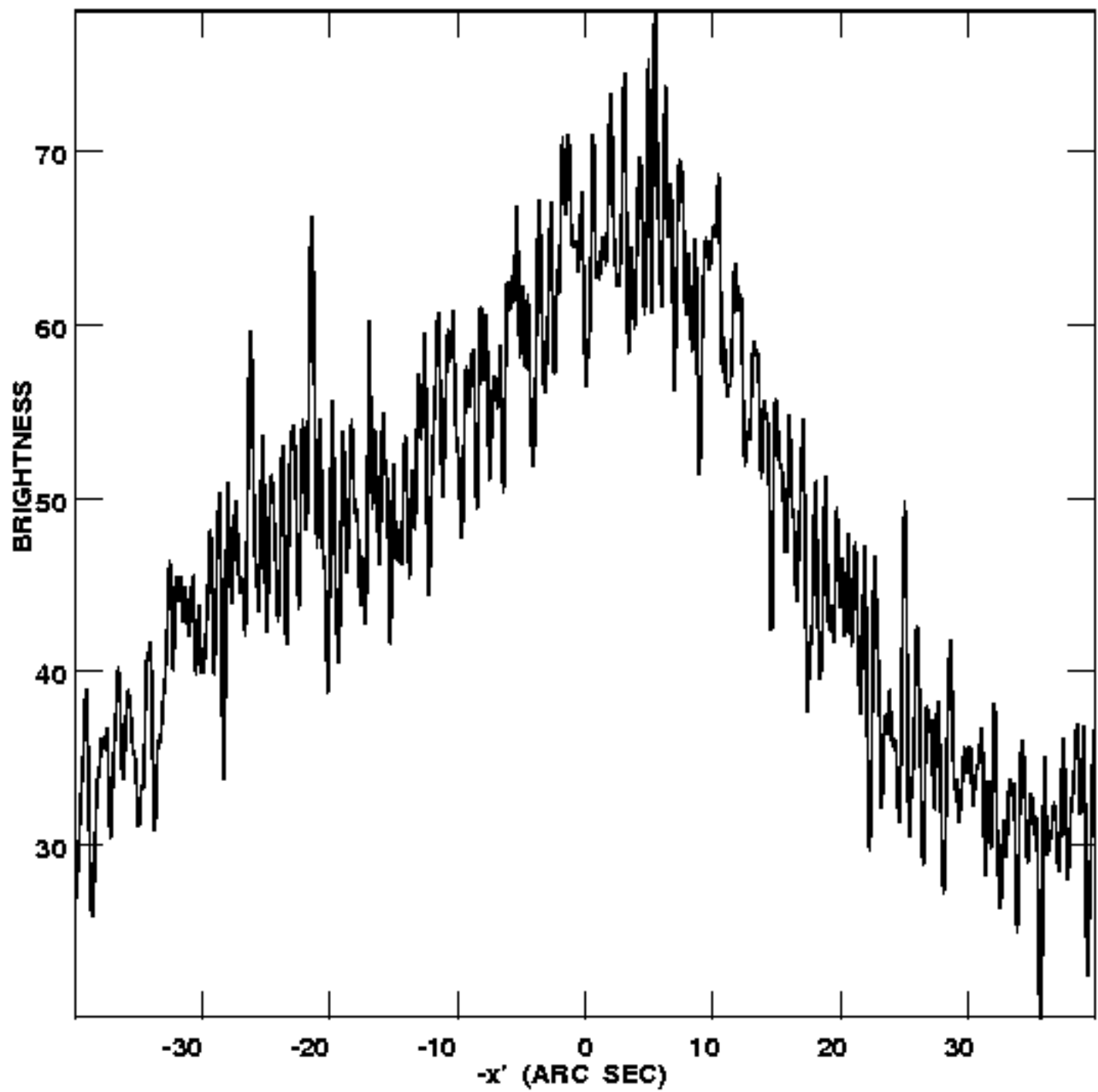
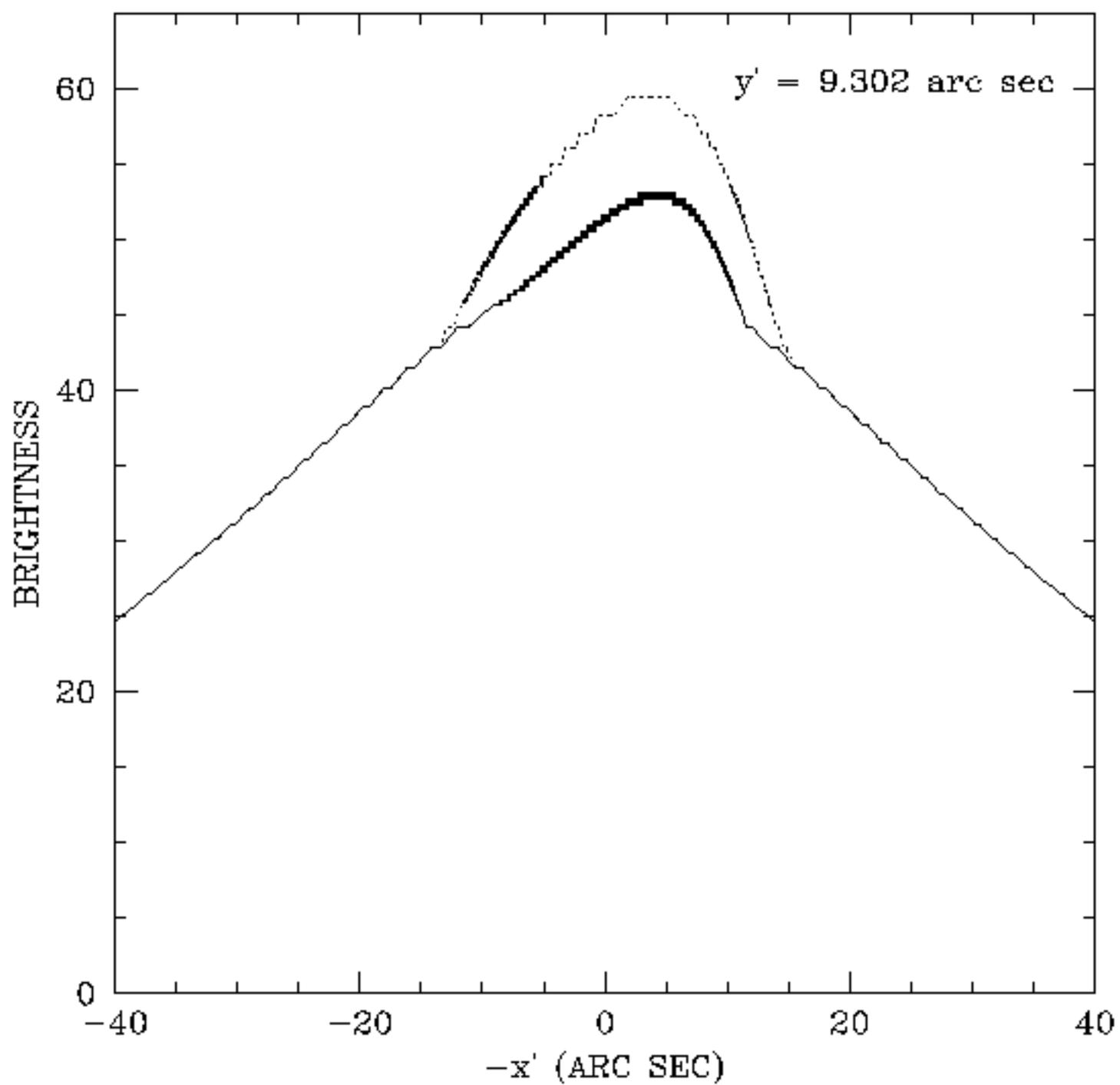
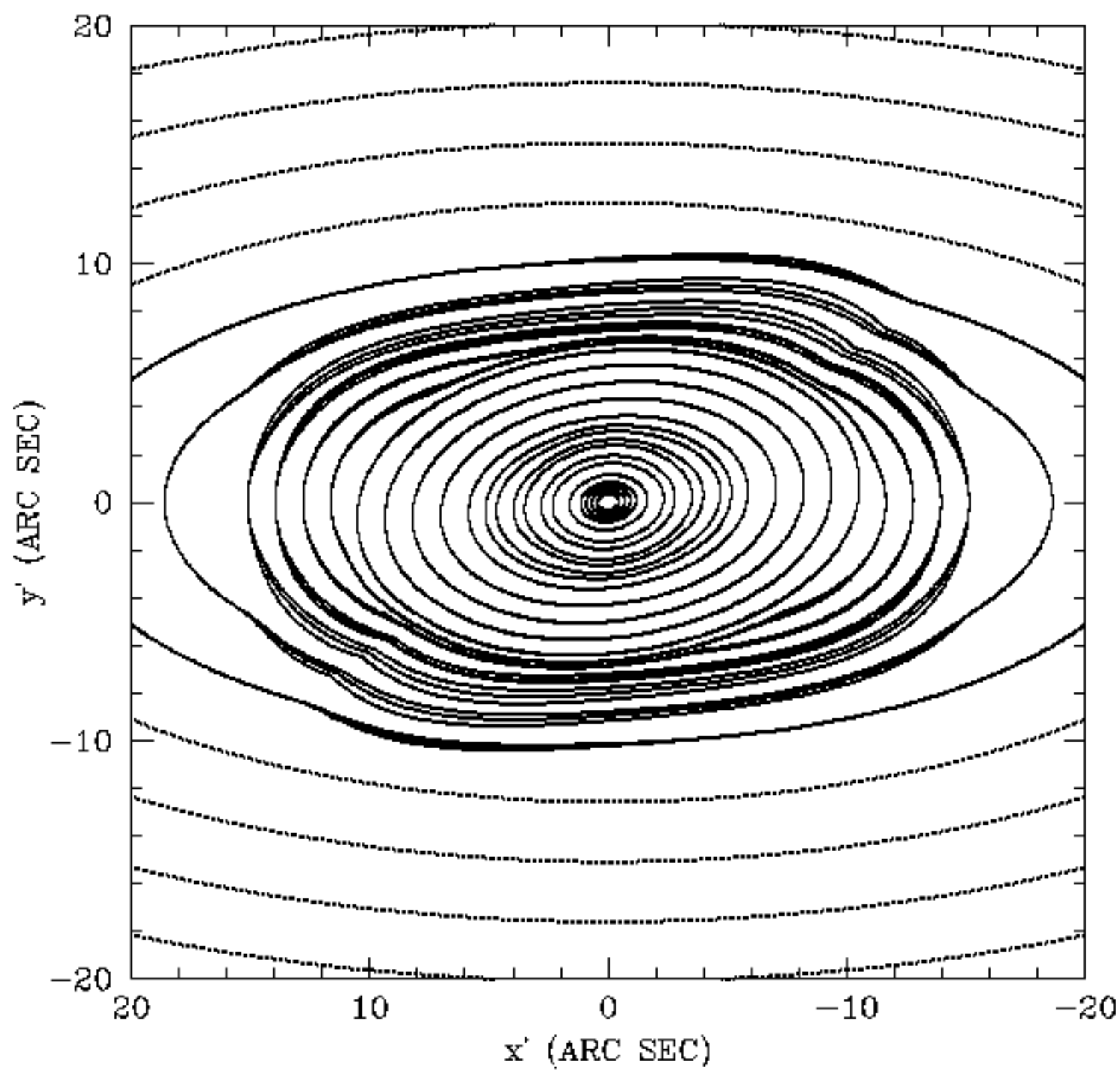
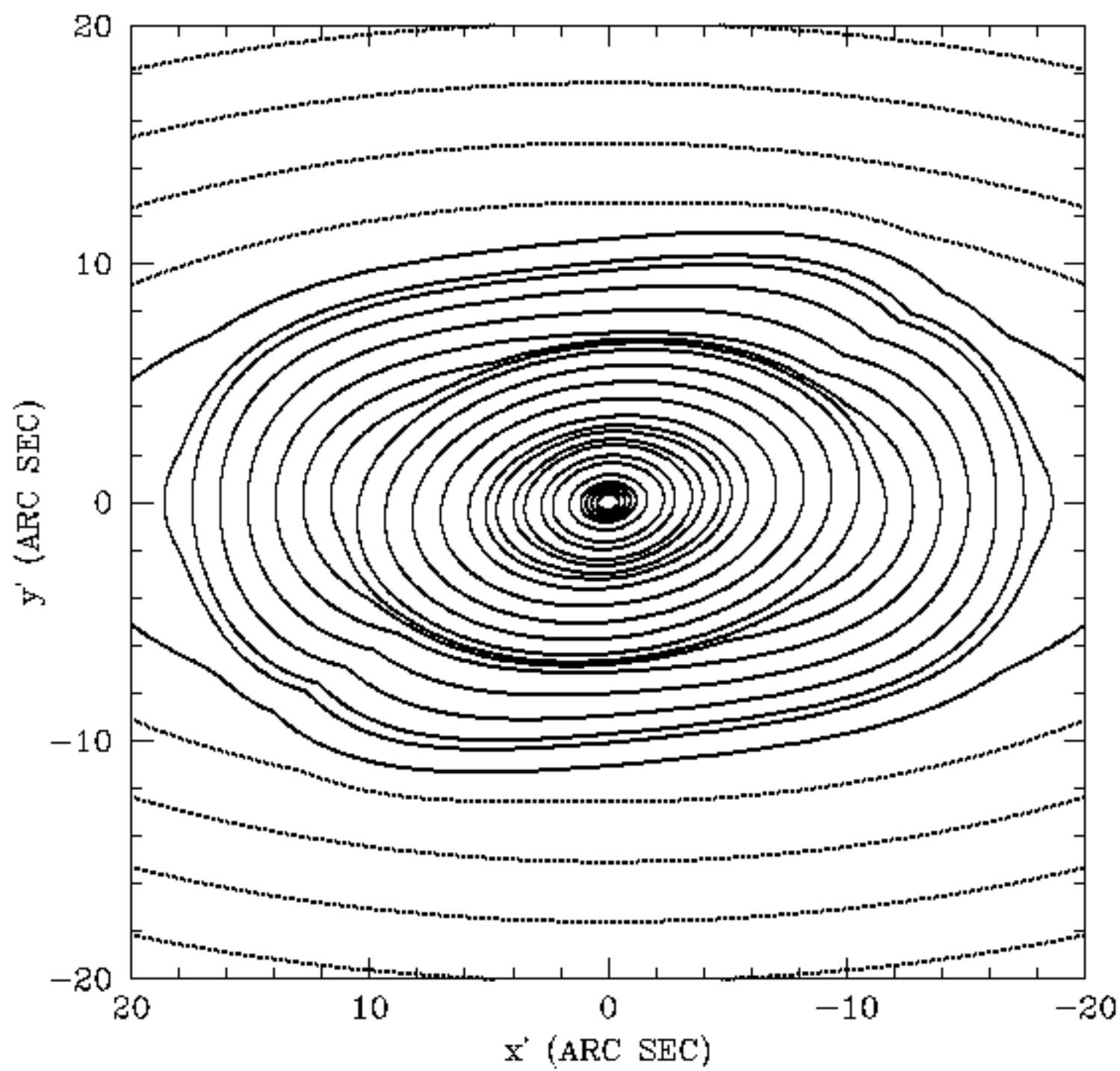
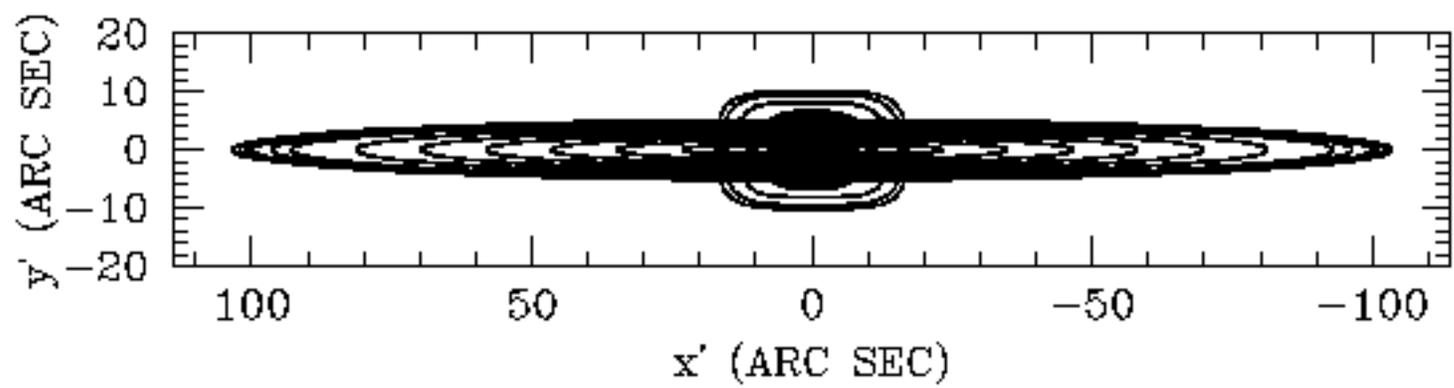
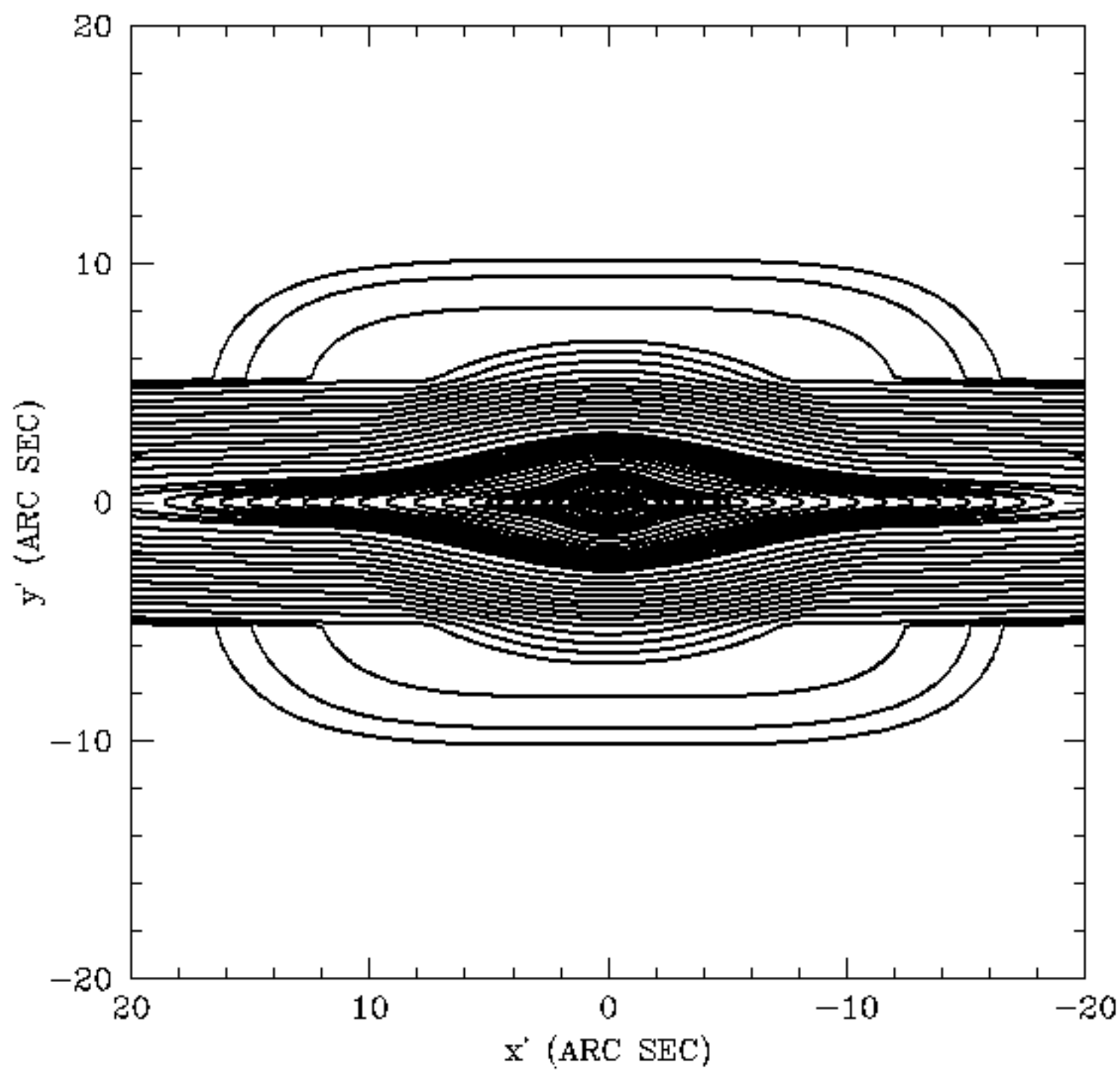
\end{document}